\documentclass[sigconf, nonacm]{acmart}

\usepackage{fancybox,framed}

\usepackage{balance}
\usepackage{makecell}
\usepackage{mathtools}
\usepackage{subfigure}
\usepackage[ruled,linesnumbered,noend]{algorithm2e}
\usepackage{pgfplots}
\usepackage{adjustbox}
\usepackage{booktabs}
\usepackage{caption}
\usepackage{graphicx}
\usepackage{amsmath}
\usepackage{soul}
\DeclareMathOperator*{\argmax}{arg\,max}
\DeclareMathOperator*{\argmin}{arg\,min}

\usepackage{colortbl}
\usepackage{url}
\usepackage{footnote}
\usepackage{tikz}
\makesavenoteenv{table}
\usepackage{pgfplotstable}
\usepgfplotslibrary{groupplots}
\usepackage{background}
\backgroundsetup{contents=Preprint}
\begin{document}

\newcommand{\ziawasch}[1]{\textcolor{purple}{Ziawasch: #1}}
\newcommand{\za}[1]{\textcolor{purple}{ZA: #1}}
\newcommand{\ma}[1]{\textcolor{cyan}{Mahdi: #1}}
\newcommand{\maedit}[1]{\textcolor{blue}{#1}}
\newcommand{\jq}[1]{\textcolor{olive}{JQ: #1}}

\newcommand{\system}{\textsc{Mate}\xspace}
\newcommand{\hash}{\textsc{Xash}\xspace}
\newcommand{\tightlist}{\itemsep=-2pt}
\newcommand*{\rom}[1]{\expandafter\@slowromancap\romannumeral #1@}

\newcommand{\Cross}{\mathbin{\tikz [x=2.5ex,y=2.5ex,line width=.5ex] \draw (0,0) -- (1,1) (0,1) -- (1,0);}}

\newcommand{\answerRone}[1]{{\color{red} {#1}}}
\newcommand{\answerRtwo}[1]{{\color{blue} {#1}}}
\newcommand{\answerRthree}[1]{{\color{purple} {#1}}}
\newcommand{\answer}[1]{{\color{orange} {#1}}}
\DeclarePairedDelimiter\ceil{\lceil}{\rceil}
\DeclarePairedDelimiter\floor{\lfloor}{\rfloor}

\newcounter{enum}
\newenvironment{packed_enum}{
\begin{list}{\textbf{(\arabic{enum})}}{
  \setlength{\itemsep}{0pt}
  \setlength{\parskip}{0pt}
  \setlength{\labelwidth}{-5 pt}
  \setlength{\leftmargin}{0 pt}
  \setlength{\itemindent}{0pt}
  \usecounter{enum}}
}{\end{list}}

\title{MATE: Multi-Attribute Table Extraction}
\author{Mahdi Esmailoghli}
\affiliation{%
  \institution{Leibniz Universität Hannover \& L3S Research Center}
  \city{Hannover}
  \country{Germany}
}
\email{esmailoghli@dbs.uni-hannover.de}

\author{Jorge-Arnulfo Quiané-Ruiz}
\affiliation{%
  \institution{TU Berlin}
  \city{Berlin}
  \country{Germany}
}
\email{jorge.quiane@tu-berlin.de}

\author{Ziawasch Abedjan}
\affiliation{%
  \institution{Leibniz Universität Hannover \& L3S Research Center}
  \city{Hannover}
  \country{Germany}
}
\email{abedjan@dbs.uni-hannover.de}

\renewcommand{\shortauthors}{}

%!TEX root = ../main.tex
\begin{abstract}
A core operation in data discovery is to find joinable tables for a given table.
Real-world tables include both unary and n-ary join keys. However, existing table discovery systems are optimized for unary joins and are ineffective and slow in the existence of n-ary keys.
In this paper, we introduce \system, a table discovery system that leverages a novel hash-based index that enables n-ary join discovery through a space-efficient super key.
We design a filtering layer that uses a novel hash, \hash. This hash function encodes the syntactic features of all column values and aggregates them into a super key, which allows the system to efficiently prune tables with non-joinable rows.
Our join discovery system is able to prune up to $1000x$ more false positives and leads to over $60x$ faster table discovery in comparison to state-of-the-art.

\end{abstract}

\maketitle

% \pagestyle{\vldbpagestyle}
% \begingroup\small\noindent\raggedright\textbf{PVLDB Reference Format:}\\
% \vldbauthors. \vldbtitle. PVLDB, \vldbvolume(\vldbissue): \vldbpages, \vldbyear.\\
% \href{https://doi.org/\vldbdoi}{doi:\vldbdoi}
% \endgroup
% \begingroup
% \renewcommand\thefootnote{}\footnote{\noindent
% This work is licensed under the Creative Commons BY-NC-ND 4.0 International License. Visit \url{https://creativecommons.org/licenses/by-nc-nd/4.0/} to view a copy of this license. For any use beyond those covered by this license, obtain permission by emailing \href{mailto:info@vldb.org}{info@vldb.org}. Copyright is held by the owner/author(s). Publication rights licensed to the VLDB Endowment. \\
% \raggedright Proceedings of the VLDB Endowment, Vol. \vldbvolume, No. \vldbissue\ %
% ISSN 2150-8097. \\
% \href{https://doi.org/\vldbdoi}{doi:\vldbdoi} \\
% }\addtocounter{footnote}{-1}\endgroup
% %%% VLDB block end %%%

% %%% do not modify the following VLDB block %%
% %%% VLDB block start %%%
% \ifdefempty{\vldbavailabilityurl}{}{
% \vspace{.3cm}
% \begingroup\small\noindent\raggedright\textbf{PVLDB Artifact Availability:}\\
% The source code, data, and/or other artifacts have been made available at \url{\vldbavailabilityurl}.
% \endgroup

\section{Introduction} \label{sec:introduction}
There is an increasing interest in enriching existing datasets with additional relevant datasets from large data silos and data lakes.
In this context, efficient data discovery systems~\cite{fernandez2018aurum,miller2018open} become essential for many use cases, such as
feature extraction~\cite{nargesian2018table, esmailoghli2021cocoa, chepurko13arda, santos2021correlation}, data cleaning~\cite{yu2016string}, 
transformation discovery~\cite{DBLP:conf/btw/ozmenEA21, abedjan2015dataxformer}, and data ecosystems~\cite{agora}.

Given a dataset at hand, the most basic requirement of a table to be relevant for enrichment is its joinability~\cite{nargesian2020organizing, xiao2009top, zhu2019josie, chaudhuri2006primitive, yu2016string, fernandez2019lazo, esmailoghlicafe, esmailoghli2021cocoa, dong2021efficient, fernandez2018aurum, DBLP:journals/pvldb/ZhuNPM16, santos2021correlation}.
However, existing works on indexing and retrieving joinable tables from large corpora are limited to unary joins, i.e.,~where a single column is the join key. 
This limitation restricts the application and the efficiency of discovery systems for prevalent datasets with composite join keys,i.e.,~join keys composed of multiple columns~\cite{jiang2019holistic}.
For example, up to $37.5\%$ of the primary keys in the TPC-E and TPC-H benchmarks are composite keys.
Further, TPC-H and TPC-E benchmarks contain over $168M$ unique column combinations (UCCs)~\cite{heise2013scalable}, $99.9\%$ of which are  multi-column UCCs. The existence of such combinations is apriori unknown and unexpected for potential discovery tasks. In open data lakes primary key information and other metadata are generally not known. Most keys are often auto-generated columns that are not directly related to the table rows and their corresponding real-world entities. 
Therefore, a discovery task has to rely on undocumented key candidates. One could discover and index the UCCs, which is an exponentially expensive task w.r.t. both runtime and storage. Considering, our next example, we cannot even claim that join columns in web tables must have the uniqueness property.

Consider the following real-world example, where we explained the root cause of air  pollution measured in different European cities\footnote{\url{https://luftdaten.info/}}.
As the sensor data is limited to the three columns (timestamp, location, and pollution ratio)~\cite{meyerparticulate}., additional dimension tables on weather, public events, and road traffics, were needed to make sense of it. All of the relevant tables had to be discovered and joined based on the timestamp and location columns of the sensor dataset~\cite{meyerparticulate}

Using current single-column discovery systems, users have to either repeat the discovery process per key column or check the retrieved joinable tables to remove the false positive (FP) tables and rows, which is a daunting task.
According to our benchmark, users might encounter over $1000$ times more irrelevant table rows.
Thus, one would first find all tables that join with the timestamp and then filter those that have the correct location or vice versa.
In either case, there would be an overhead for removing irrelevant tables.
For example, searching for joinable tables in the Dresden Web Table Corpus\footnote{\url{https://wwwdb.inf.tu-dresden.de/misc/dwtc/}}) based on the location and timestamp leads to $700K$ and $4M$ candidate tables, respectively.
However, only $1,552$ of them contain the exact alignment of time and location that is useful to discover the pollution reasons.

Using a state-of-the-art system, such as JOSIE~\cite{zhu2019josie}, the runtime for verifying both columns for joins increases by an order of magnitude because discovering tables based on n-ary keys has a factorial time complexity in the number of attributes in each table. Besides, in contrast to the single column key scenario more candidate tables have to be scanned to identify the top-$k$ joinable tables as it is not guaranteed that the joinability of each join column is equally high in each candidate table. 
Leveraging an inverted index to achieve the efficient discovery of joinable tables is crucial, but building a multi-attribute inverted index requires factorial storage space, which is infeasible.

We propose \system, a data discovery system to efficiently find n-ary joinable tables from a large corpus with millions of tables for a given query table.
To the best of our knowledge, this is the first piece of work addressing the general dataset discovery problem for n-ary key joins at a large scale.
\system employs a pruning technique to detect the most promising table rows and apriori filter irrelevant tables.
It uses a new inverted index element called \emph{super key}, which is an order-independent and fixed-sized hash-value that merges all possible key values into a single index element.
This super key allows the system to check the existence of any given key value in the same spirit as a bloom filter.
To generate the super key, \system uses \hash a simple, yet effective, h\textbf{ASH} function for cheking the e\textbf{X}istance of arbitrary key values. It encodes all cell values based on distinctive properties to avoid collisions of similar join values. As a result, it significantly reduces the FP rate of joinable rows before searching for join attributes and exact matches within a row. 
% Let us illustrate this with an example: consider a data scientist that aims to discover the correlating features to the IMDB score of movies. She would like to explain why the \textit{director} and the \textit{actors} involved in movies impact the IMDB score and how. The key in this example could be the \textit{movie title}, where the name of the movie defines all the relevant information including the \textit{director} and the \textit{actors}. However, joining tables based on the \textit{movie title} will not help with the task at hand. The data scientist needs to find additional features about the \textit{actors} and the \textit{director} of the movie and see if the integrated columns can lead to a reasonable prediction power.
In addition to multi-column join discovery, other use cases and application can directly benefit from our approach to beat the exponential dimensionality of multi-column sets. The methods are readily adaptable for duplicate table discovery and union table discovery~\cite{lehmberg2017stitching, DBLP:journals/pvldb/CafarellaHK09, das2012finding, nargesian2018table, DBLP:conf/icde/BogatuFP020}, and spreadsheet transformation joins~\cite{DBLP:journals/pvldb/ZhuHC17}. For duplicate table detection, our hash function could serve as a prefilter for finding similar records. For table union search, the hash function could be applied in the same spirit as for joins.

In summary, our major contributions are as follows: 
% \vspace{-0.1cm}
\begin{packed_enum}
\item We formulate the general problem of table discovery with n-ary key joins and take the first step in solving this problem with a filter-based approach. 
\item We introduce \hash, which encodes the syntactic attributes of the key values to obtain hash results that optimally use a fixed bit space to avoid overlapping bits of similar join values. The hash function leads effectively filters non-joinable rows and fewer FP rates compared to the state-of-the-art hash functions and bloom filters.
Our super key simulates a multi-attribute inverted index with a fixed size hash.
We prove that our hash function does not cause any false negatives.
\item We introduce a two-tier filtering strategy that apriori prunes candidate tables that cannot be part of top-$k$ and candidate rows that are not joinable on all key columns without evaluating the actual row values.
\end{packed_enum}

%We conduct extensive experiments on two large corpora of open data datasets to evaluate our system. According to the experiments, our hash-based super key index leads to
%up to almost $20$ times faster discovery compared to the existing baseline. 
%The hash function used in \system is able to prune almost $500$ times more irrelevant rows than other state-of-the-art hash functions.

%!TEX root = ../main.tex

\section{Problem Statement}\label{sec:problem_statement}

We focus on discovering joinable tables based on composite key joins.
Generally speaking, the problem of table discovery is to find the top-k joinable tables for a given query table with a selected composite key~\cite{zhu2019josie}.
We first formalize the joinability between two tables and then define the problem of n-ary join discovery from large data lakes.
We borrow the notations from the literature on inclusion dependencies~\cite{papenbrock2015divide, de2009unary}.

\noindent\textbf{Joinability.} 
Intuitively, the joinability between a candidate table (from a corpus) and a given query table represents the corresponding equi-join cardinality.
That is, the more key values in a candidate table can be joined with the query table the higher their joinability.
Thus, joinability is a measure for the completeness of the join and the relevance of a candidate table to a query table.
Formally, assume that $\mathcal{R}$ and $\mathcal{S}$ are two relational schemata and attribute sets $X$ and $Y$ are two subsets of columns where, $X \subseteq \mathcal{R}$ and $Y \subseteq \mathcal{S}$. Without loss of generalization, we can pick any $X$ and $Y$ that consist of the same number of attributes: $|X| = |Y| = m$.
In multi-attribute joins $m > 1$. If $r$ is a set of tuples over $\mathcal{R}$, the projection of $\mathcal{R}$ onto $X$ is shown by $\pi_X(\mathcal{R})$, where $\pi_X(\mathcal{R}) = \{t[X]|t \in r\}$. 
Likewise, $s$ is a set of rows over $\mathcal{S}$.
We, thus, define the joinability score $\jmath$ between $\mathcal{R}$ and $\mathcal{S}$ on the selected column sets $X$ and $Y$ as:
\begin{equation}\label{eq:joinability}
\small
    \jmath(\mathcal{R},\ \mathcal{S}) = |\pi_X(\mathcal{R}) \cap \pi_Y(\mathcal{S})|.
\end{equation}
Yet, calculating Equation~\ref{eq:joinability} is not possible because the one-to-one mapping between the key columns in $\mathcal{R}$ and $\mathcal{S}$ is unknown.
Any column permutation of size $|X|$ is a possible candidate.
Thus, the joinability between a table with a given composite join key and a table without a defined join key is a factorial number of possible mappings in the number of join key columns.
This is why we extend the joinability definition as follows:
\begin{equation}
\small
    \jmath(\mathcal{R},\ \mathcal{S}) = \argmax_{Y'} {|\pi_X(\mathcal{R}) \cap \pi_{Y'}(\mathcal{S})|}.
\end{equation}
Here, $Y'$ is a permutation of size $|X|$ from $\mathcal{S}$  where $\jmath(\mathcal{R},\ \mathcal{S})$ is maximum.

\begin{figure}
    \center{\includegraphics[scale=0.14]
          {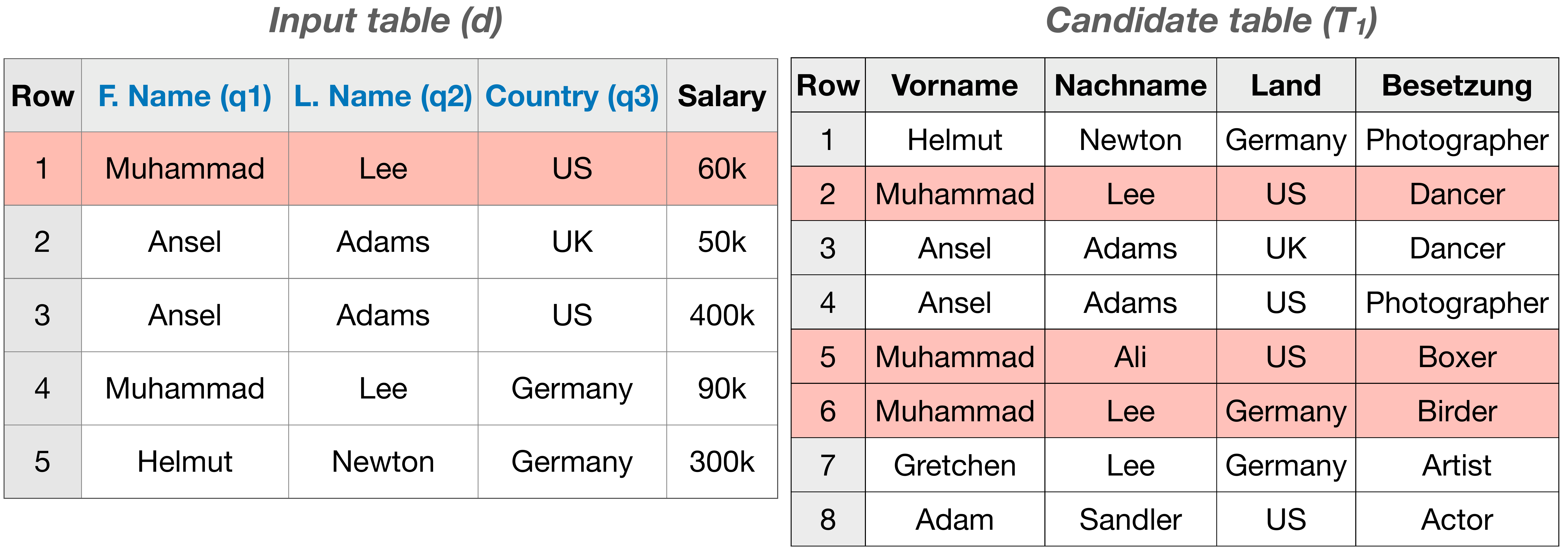}}
% \vspace{-.5cm}
	\caption{Running example.}
    \label{fig:example}
\end{figure}

% \vspace{0.1cm}
\noindent{\em Running example.}
Consider a query table $d$ and a candidate table $T_1$ as illustrated in Figure~\ref{fig:example}.
Assume the user has selected \textit{F. Name}, \textit{L. Name}, and \textit{Country} as the query columns (columns with blue header).
We aim at finding the three columns from table $T_1$ that result in the highest joinability ($\jmath$) with the query columns from $d$.
If we map \textit{F. Name} from $d$ to \textit{Nachname} from $T_1$, map \textit{L. Name} to \textit{Vorname}, and map \textit{Country} to \textit{Land}, the one-to-one mapping would lead to a $\jmath$ of $0$;
If we map \textit{F. Name} from $d$ to \textit{Vorname} from $T_1$, map \textit{L. Name} to \textit{Nachname}, and map \textit{Country} to \textit{Land}, we would obtain $\jmath=5$, the maximum joinability score among all possible mappings.

% \vspace{0.1cm}
\noindent\textbf{The n-ary join discovery problem.} Given a base table $d$ with a relation $D$,
a set of query columns $Q$, where $Q \subset D$, a corpus of tables $T$, and a constant value $k$, the goal is to return the top-$k$ tables from $T$ sorted by their joinability $\jmath$.
Solving this problem is challenging for two main reasons:
\begin{packed_enum}
    \item Calculating the joinability between a given query table and a candidate table with a set of attribute $T'$ requires the mapping between $Q$ and $T'$ that maximizes $\jmath$.
    The number of possible mappings is calculated as:
        \begin{equation}
        \small
           P(|T'|, |Q|) = \frac{|T'|!}{(|T'| - |Q|)! |Q|!}
        \end{equation}
        Here, $P(|T'|,\ |Q|)$ represents the number of possible permutations with size $|Q|$ out of $|T'|$ columns.
    \item Ranking tables based on $\jmath$ and picking the top-$k$ requires calculating the joinability on each candidate table in $T$. 
\end{packed_enum}

To discover n-ary joinable tables, one has to (i)~build a multi-attribute inverted index that maps different combinations of cell values to their locations in the tables, or (ii)~use the inverted index for unary joins and tolerate a large number of FPs.
Building a multi-attribute inverted index is not feasible with respect to the storage complexity.
For each of the 145M tables inside the Dresden Webtable Corpus, one would need to create $\sum_{i=1}^{c} P(c,\ i)$ indexes per table.
For $c=5$, the size of the database would increase by more than one order of magnitude.
We propose a filtering solution that extends the single-attribute inverted index to obtain both time and space efficiency.

\section{Preliminaries}\label{sec:preliminaries} 
Our approach for enabling multi-attribute joins extends the common inverted index structure.

\textbf{Inverted index.} The inverted index is a structure that maps the content, such as tokens or words, to their containing structures, i.e., tables, rows, and columns~\cite{fernandez2018aurum, abedjan2015dataxformer}. In this work, we extend the index as proposed for the DataXformer system~\cite{abedjan2015dataxformer}: 
\newline
\begin{equation} \label{eq_conventional_inverted_index}
% \vspace{-.2cm}
\small
    v_i \mapsto PL_i = \{ (T_{i1},\ C_{i1},\ R_{i1}),\ (T_{i2},\ C_{i2},\ R_{i2}),\ ... \}.
\end{equation}
% \vspace{-.2cm}
\newline
where, $v_{i}$ is a value and $T_{ij}$, $C_{ij}$, and $R_{ij}$ are the identifiers of the corresponding tables, columns, and rows in the corpus, respectively. This list of triplets is called \textit{Posting List} (PL). We also call every single triplet a PL item.

\textbf{Discussion. } 
Many state-of-the-art systems leverage the inverted index with small alternations~\cite{zhu2019josie,xiao2009top,chaudhuri2006primitive}.
Given a single-attribute join key, the current systems retrieve the PL items for each value in the key column and the number of returned PL items represents the joinability score for each table.
To use the same benefits and optimizations for n-ary joins, we need to generate a multi-attribute inverted index that maps every possible combination of cell values to their location in the tables. For instance, in our running example, we would like to have a PL item that maps the key value of <``Muhammad'', ``Lee'', ``US''> to the rows that contain both of these values at the same time.

One can use a straightforward algorithm that leverages the original inverted index to find the multi-attribute joinable tables. This algorithm obtains the PLs of one single query column first and then verifies whether the values of the remaining query columns appear in the same tables and rows.

As previously discussed, this approach leads to a large number of false positive rows that require a second verification step.
A \textit{false positive row (FP row)} is a row from a candidate table that only contains a subset of join attribute values. For instance, if the search goal is to find joinable tables based on a given 2-column key, candidate rows that are retrieved based on one attribute of the join key and thus only contain one value of the key value combination are considered as FPs and should be excluded from the joinability calculation.
A \textit{false positive table (FP table)} is a candidate table with FP rows that is not among the top-$k$ joinable tables. From here on, we refer to FP rows as FPs unless we explicitly specify the type of FPs.
The FP rate can be up to $1000$ times higher than the actual number of joinable table rows. For each additional row, the discovery system has to compare each value to the composite key values of the input query.

\textit{Example 2.} Going back to our example in Figure~\ref{fig:example}, for the value ``Muhammad'' in the $1^{st}$ row of the query column $q_1$, there are three hits in the $2^{nd}$, $5^{th}$ and $6^{th}$ rows of $T_1$ and column \textit{Vorname}. These rows are highlighted in red. To find the exact matches for the remaining key values ``Lee'' and ``US'' in the second and third query columns ($q_2$ and $q_3$), the system has to check every other value in rows $2$, $5$, and $6$ in Table~$T_1$. Here, \system checks $9$ value matches only for the first key value <``Muhammad'', ``Lee'', ``US''>.
Alternatively, one would have to send two other independent queries against the index to obtain all rows where ``Lee'' and ``US'' occur. 
However, if an oracle could confirm or deny the existence of ``Lee'' and ``US'' with a single operation, the system would only need to check $3$ values.
This optimization would drastically improve the runtime for calculating the joinability score $\jmath$ for multi-attribute keys.
Thus, an index element is required that conveys evidence of whether a particular row contains a given composite key or not. We call this additional element, \textit{super key}. 
%In the next section, we explain our system design and we show how \system generates the \textit{super key}.

%!TEX root = ../main.tex

\section{System Overview} \label{sec:system}
\begin{figure}
% \hspace{-.8cm}
    \center{\includegraphics[scale=0.19]
          {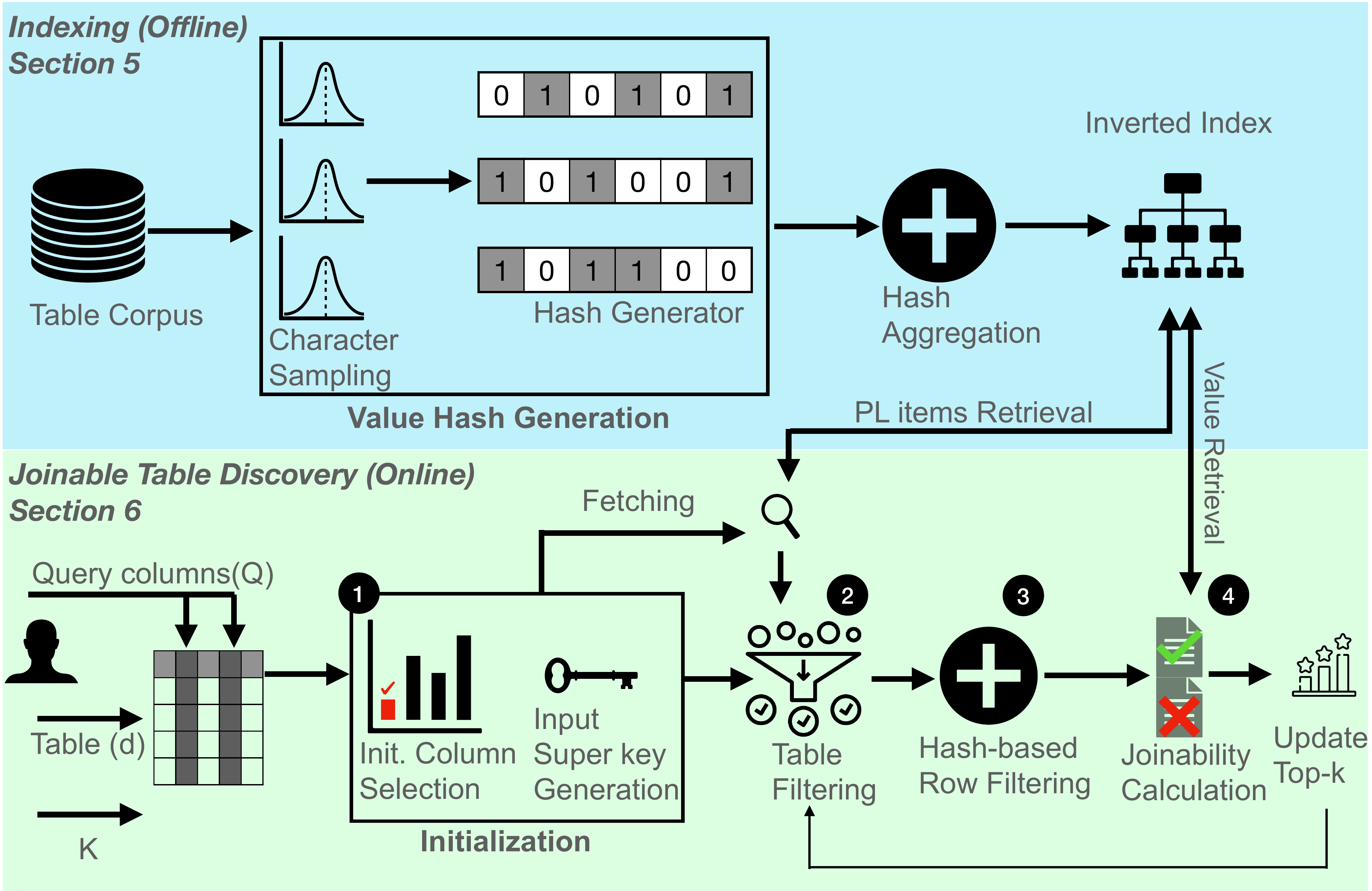}}
            %   \vspace{-.2cm}
    \caption{The overall workflow of \system.}
    \label{fig:architecture}
\end{figure}
Figure~\ref{fig:architecture} depicts the abstract workflow of our proposed system, \system. %It consists of two main phases: the {\em offline} phase and the {\em online} phase.
In the offline phase, \system builds an inverted index structure for the tables in the corpus. 
In the online phase, i.e. discovery phase, \system uses the inverted index to find the top-k joinable tables among all candidate tables for a given input table. 

% \vspace{0.1cm}
\noindent\textbf{Indexing step (Offline).} To efficiently discover tables that are joinable to a given table, we propose an extension to the state-of-the-art single-value inverted index to efficiently serve the multi-attribute applications.
We introduce an additional index element, which is an aggregated hash value called \emph{super key}.
The \emph{super key} is space-efficient and does not change the nature of the single-attribute inverted index while serving the purpose of multi-attribute join discovery.
The super key is used to prune irrelevant tables and rows to reduce the post-processing overhead. It aggregates the values inside each table-row into a fixed-size entry.

It is worth noting that the super key element does not require any knowledge of the actual keys of a table and serves all possible key combinations inside a table.
This way, for a given query dataset and composite key, the system can decide in a single operation if the row is a candidate joinable row or not.
Furthermore, the filter does not result in any false negatives.
To generate the super key entries, \system leverages a novel hash function \hash that encodes each row value in a highly disseminative way and aggregates them to the super key. The hash function, as well as the super key, will be discussed in more detail in Section~\ref{sec:index}.

% \vspace{0.1cm}
\noindent\textbf{Discovery step (Online).} In the actual data discovery scenario, the user provides a dataset with a composite key and a parameter $k$, expecting the system to find top-$k$ tables with the highest number of equi-joinable rows to the given input dataset.
\system enables the efficient n-ary key joins by using the super key entry to prune as many irrelevant tables as possible before the joinability calculation.
This process undergoes four phases: \textit{(i)}~initialization, \textit{(ii)}~table filtering, \textit{(iii)}~ row filtering, and \textit{(iv)}~the joinability calculation.

In the initialization step, first, the system selects an initial query column from the composite key \textit{Q} based on a simple cardinality-based heuristic. The goal of the initial query column is to reduce the number of fetched tables to the minimum by picking the column that leads to the smaller number of PL items from the corpus. 
That is why \system selects a single column to fetch the initial set of candidate tables from the corpus.
The column selection can be supervised and preempted by the user.
Then, the system generates the super key entry for the join columns of the input dataset, i.e.,~generating a hash-code for each key value combination and aggregating them into a single hash. 
In the filtering step, \system applies two levels of pruning for each table: table-level and row-level. 
With the table-level pruning, \system decides whether the current table is a promising table to be one of the top-k tables.
With the row-level pruning, \system checks for non-joinable rows in the candidate table. It compares the super keys of the input dataset with those inside the candidate table to drop irrelevant rows from further joinability verification.
Finally, \system retrieves the exact values from the table corpus to compute the final joinability score $\jmath$ for the remainig tables and rows. After calculating the $\jmath$, the set of top-$k$ tables is updated and the next candidate table undergoes the pruning steps.
We detail the online phase in Section~\ref{sec:index_applicaiton}.

%!TEX root = ../main.tex

\section{Indexing and Super Key Generation}\label{sec:index}
We propose an index structure that retains the space efficiency of the single-attribute inverted index as introduced in Section~\ref{sec:preliminaries}.
% Table~\ref{tab:variables} summarizes the variables used in this paper.

% \input{charts_and_tables/variables}

\subsection{Desiderata}
\label{sec:desiderata}
As generating a multi-attribute inverted index requires an exponential number of index entries per table,
it is important to extend the traditional single-attribute inverted index to be applicable for n-ary joins.
Ideally, we need an additional index element per table row that exposes the existence of join attribute value combinations.
We, thus, add an additional element to the index structure named \textit{super key}. This changes the inverted index defined in Equation~\ref{eq_conventional_inverted_index} to $v_i \mapsto \{(T_{i1},\ C_{i1},\ R_{i1},\ S_{i1}), (T_{i2},\ C_{i2},\ R_{i2},\ S_{i2}),\ ... \}$, where $S_{ij}$ is a fixed-size bit array, i.e.,~\textit{Super Key}.

As one cannot anticipate which attribute combinations are relevant for n-ary joins, we need a hash value that retains them all.
The idea is to have a fixed size super key that aggregates hash values for each row value so that we can verify the existence of a composite key without checking all values of the row.
\system aggregates the hash results of individual values into a fixed-sized bit vector using the logical bit-wise \texttt{OR} operation. Thus, the super key masks the hash values of each single row value, which ensures that no value is missed, when probing with the super key with the same hash function.

Consider our previous example once again. Rows $2$, $5$, and $6$ are candidate rows to be joined with the first key in the input dataset d based on the first value ``Muhammad''. Now, to drop the $5^{th}$ and $6^{th}$ row from the candidate rows, the super key for these rows should convey that the values ``Lee'' and ``US'' do not simultaneously exist in these candidate rows.
Yet, the drawback of the aggregated hash value within a fixed hash table is that the super key might mask the hash values of non-existent values as well.

Therefore, the goal is to design a hash function where the aggregation of cell values from different columns results in different super keys.
A general approach to this problem is to use a bloom filter.
However, off-the-shelf bloom filters have two drawbacks.
First, they use hash functions that assume a uniform distribution.
Therefore, any arbitrary pair of cell values can result in overlapping bits in the final bit array, which increases the chance of FPs.
Second, they are agnostic to the distinguishing properties within columns, i.e., character distributions and positions.

%\vspace{0.1cm}
%\noindent{\bf Desired Hash Function.} To overcome the aforementioned shortcomings of bloom filters, we need a hash function whose parameters are independent of the query table and the individual tables of the corpus. The {\em desired hash function} should encode values in a way that syntactically similar values in different columns and syntactically different values of the same column fall into different bit positions of the super key. As the hash space and ultimately the number of bits are limited, we can only hash a finite number of value features, which have to be the most differentiating properties of a row value.

\subsection{\hash}\label{subsec:synhash}
We propose \hash, a hash function that encodes syntactic features into distinguishable hashes.
As the super key is an OR-aggregation of these hash results, it may mask non-existent values and pass FPs. 
Thus, our goal is to disperse 1-bits in a way that we decrease the likelihood of two different values from different columns turning the same set of bits to 1. Ideally we want as few 1-bits as possible in the super key to reduce the probability that it covers the super key of random value combinations.
\hash leverages three syntactic properties of the cell values to meet this goal:
the {\em least frequent characters}, {\em their location}, and the {\em value length}.

\subsection{Feature Generation and Encoding} \label{subsec:hash_generation_process}
We now turn our attention to how we extract the aforementioned features to apply \hash on a row value.
We first discuss the number of bits to generate the hash and its relationship to the table corpus.
Then, we explain our segmentation process that allocates different parts of the hash table for different features of a value.
Then, we elaborate on the process of mapping the character features to hash bits.
Finally, we explain how to use the hash space to relocate the generated bits per cell value to prevent partial matches.
We use the example in Figure~\ref{fig:hash_array_example} to elaborate each hash generation step.

\subsubsection{Required number of bits}
The super key encodes each value into a fixed-size bit array $a$. 
On the one hand, as we want the hash values of different individual strings to differ, we want to use all possible bits to encode as many values as possible, i.e.,~$2^{|a|}$, to reduce the number of collisions.
On the other hand, the super key should contain as few `$1$s' as possible to avoid masking FPs.
As a result, underpinning \hash results should be constructed in a way that there is an upper bound of $1$ bits for each hash.
With this goal in mind, Equation~\ref{eq:number_of_ones} calculates $\alpha$, i.e., the optimum number of `$1$s' that is required to generate unique hash values, where $C_{unique}$ is the number of unique values in the corpus and $|a|$ is the hash size.
\begin{equation}\label{eq:number_of_ones}
\small
    \argmin_{\alpha} {|a| \choose \alpha} > C_{unique}.
\end{equation}
The binomial term calculates the number of possible bit combinations of size $\alpha$ over $|a|$ bits.
The minimum value for $\alpha$
corresponds to the number of `$1$' bits needed per \hash result.
For a $128$-bit hash space and $700M$ unique values as existing in DWTC webtables, $\alpha$ is equal to $6$. 
In fact, out of the $\alpha$ bits, one bit is always reserved to encode the value size and $\alpha-1$ bits for the actual value.
Assume that the hash size in our illustrating example is $128$ bits and $\alpha = 4$ ($3$ for the characters and $1$ for the value length).

\subsubsection{Encoding the characters}\label{subsubsec:encoding_chars}
As we use $\alpha-1$ bits to encode each value, we want different values to use different bit segments of $a$.
Thus the characters should be maximally different across words. We can obtain this property based on the character frequency.
\textbf{Lemma.} Least frequent characters lead to fewer collisions.
% \vspace{-.3cm}
\begin{proof}[Proof.]
Given a random word $w_1$ consisting of letters ${l_1,..l_n}$ each with a probability of occurrence $P(l_i)$, we sample $K<n$
letters $S= {s_1,..,s_k}$, where $K=\alpha - 1$. 
Given a random word $w_2$ from the same alphabet, we want to reduce the probability to sample the same set of characters $K$.
The probability to obtain a word with the same $K$ characters is $P(s_1)\cdot P(s_2)\cdot\dots\cdot P(s_k)$. 
This product is minimised whenever a factor is replaced with a smaller probability. Thus, when picking the $K$ least frequent character of the alphabet we obtain $\forall \hat{s}_i \in \{w_1 - S\}, \forall s_i\in S: P(s_i) < P(\hat{s}_i)$. 
Replacing any $s_i$ with $\hat{s}_i$ results in $P(s_1)\cdot P(s_2)\cdot\dots\cdot P(s_k) < P(s_1)\cdot P(\hat{s}_i)\cdot\dots\cdot P(s_k)$, which leads to a higher probability for a collision.
\end{proof}

To further distinguish the frequencies across domains, we pick the $\alpha-1$ least frequent characters inside a word as the differentiator. For single-word cell values with flat frequency distributions we draw based on lexicographical order of the characters.

\noindent\textit{Segmentation.}
We segment the hash space into smaller blocks, depending on the length of the hash array to encode the features of a cell value.
This segmentation specifies how many bits each feature needs. 
Depending on the number of possible characters, \hash splits the hash array into as many smaller fixed-size segments, one for each character. 
In our case, we consider all 37 alphanumeric characters including space, which results in 37 segments of size $\beta$.
We create an additional segment to encode the length of the string value.
Sticking with 37 as the number of characters, we can calculate $\beta$ as follows:
\begin{equation}
\small
    \argmax_{\beta} {(37 \cdot \beta < |a|)}
\end{equation}
$|a|$ is the length of the hash array: we have $\beta=3$ with a hash size of \textit{128} bits.
This segmentation dedicates the largest possible sub-array to encode character features, because the character features, including the position of the characters, are more discriminative than the length of the value.
The rest of the hash array can be allocated to the length segment: $|a_l| = |a| - (37 \cdot \beta)$.
For a hash size of $128$ bits, the length segment would comprise of $17$ bits ($128 - 37 \cdot 3$). 
$99.99\%$ of the English words have fewer than $17$ characters~\cite{mayzner1965tables}. Therefore, the explained segmentation can cover almost all possible words in English language. In our real-world data lakes, Dresden webtable and German open data, over $83\%$ of the cell values have at most $17$ characters. For larger hash sizes, i.e., $512$, $|a_l|=31$  and covers the length of more than $98\%$ of the cell values in our corpora.

\begin{figure}
    \center{\includegraphics[scale=0.14]
          {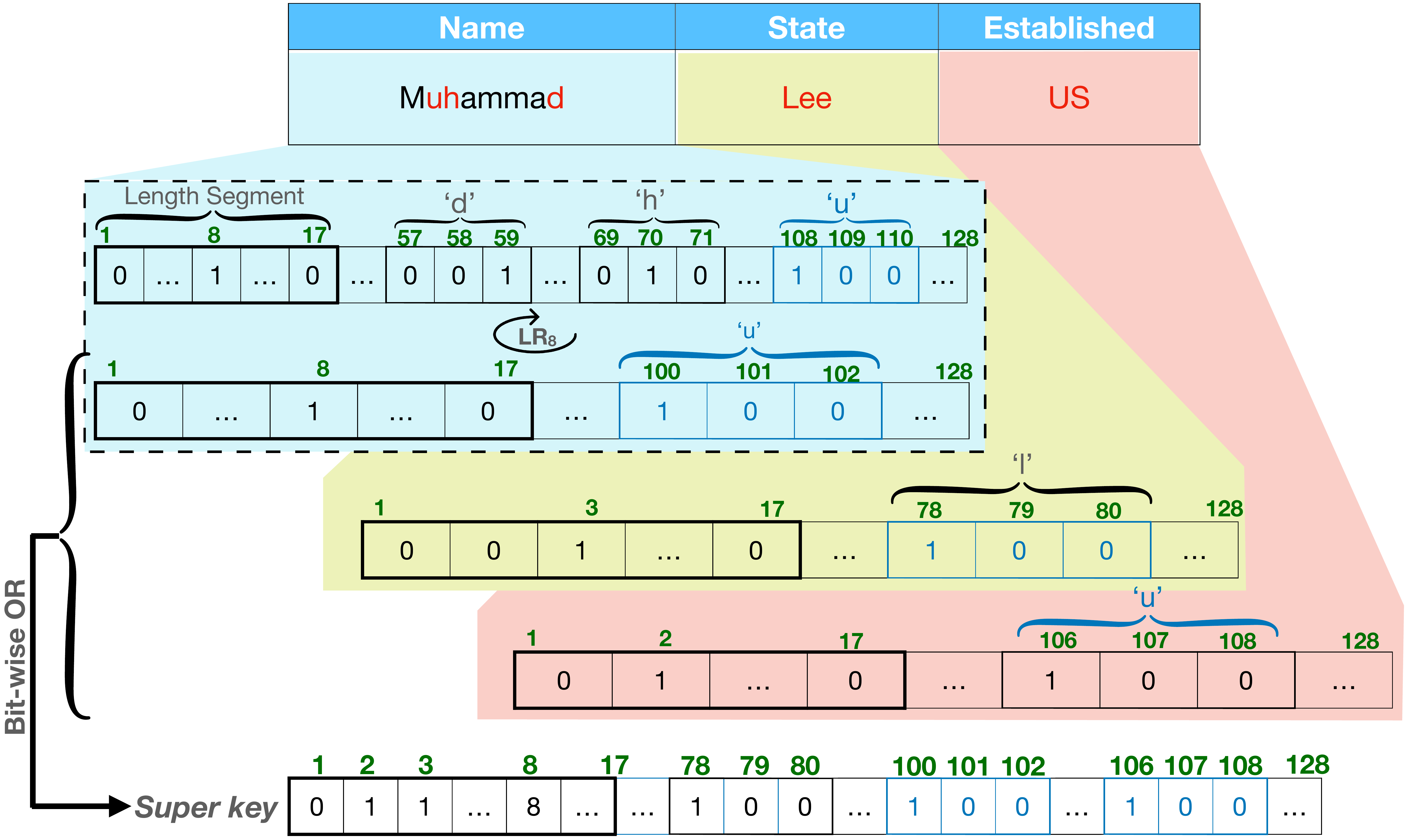}}
            %   \vspace{-.6cm}
    \caption{Example of \hash and the results aggregation.}
    \label{fig:hash_array_example}
\end{figure}

In Figure~\ref{fig:hash_array_example}, we want to obtain the hash results for values ``muhammad'' ($C_1$), ``lee'' ($C_2$), and ``us'' ($C_3$) using \hash and then aggregate them into the super key of the row. The red characters in the table cells represent the selected least-frequent characters.

\subsubsection{Encoding the character locations}
Generally, one bit per character would be enough to encode its occurrence in a value.% i.e., ``0'' for absence and ``1'' for the existence of the character.
However, if more bits are available ($\beta>1$), we can use them to encode the relative character position inside the original string to further distinguish the hash results of different values. For this purpose, we divide the string in $\beta$ equal areas from left to right. 
We encode a character by checking in which area the character appears and then we set only the corresponding bit among the $\beta$ character bits from left to right.
More formally, if $l_v$ is the length of the value, i.e., the number of characters in the value, and $\lambda$ is the location of the character we would like to encode (we take average of the locations in case of character repetition), then, $x$, i.e., the bit index in the character segment that represents the relative location of the character is calculated as: $x = \ceil*{\frac{\lambda\cdot \beta}{l_{v}}}$

% We illustrate this with an example:
% For $\beta=1$, we can only store the absence/existence of a character, therefore, the hash values would be equal for the two words ``loop'' and ``pool''. This is because both values are comprised of the same characters thus, the same character segments would be set and the length is equal. However with $\beta=3$, we assign $3$ bits per character in the hash array and consider three consecutive areas of each string. Now, character hash segments will look differently. In the hash of the value ``loop'', the left most bit of segment ``L'' turns to 1, i.e.,~$<100>$, because character ``L'' is located in the most left part of the string.
% For ``pool'', this bit assignment of character ``L'' would be $<001>$ because ``l'' appears in the last third area of ``pool''.
% The opposite applies to the encoding of ``P''.
% The encoding for ``O'' as the average of the two middle positions turns the middle bit of the ``O'' segment for both hash results.

Returning to our illustrating example, we calculate the average location in the original value for each of the least frequent characters.
For instance, the average location of the characters ``u'' and ``d'' in cell $C_1$ are $1$ and $8$ respectively.
The first hash array represents the \hash result of $C_1$.
With 3 bits for each character ($\beta = 3$):
the characters with the average location of less than $3$ ($\frac{8}{3}=3$) turn the first bit (100);
between and including $3$ and $5$ turn the second bit (010); and
above $5$ are encoded with the third bit of their corresponding segments (001).
For example, characters ``u'' and ``d'' in $C1$ are located in the first and the third bits of their segments respectively.
We use blue color to trace the character ``u'' in the hash array.

\subsubsection{Encoding the length}\label{subsub_length}
With the segmentation step of \hash, a fixed-size segment is dedicated to the length of the value $l_v$.
Storing the actual binary representation of the value length can be problematic as it leads to an unknown number of $1$ bits in the segment. This can also mask the length values of the columns with shorter values. For instance, the encoding of $l_v=7$,i.e., $<111>$, can mask the encoding of a value with $l_v=2$, i.e., $<010>$.
To address this problem, we reserve one bit per possible $l_v\mod |a_l|$. This way, we maintain the number of used bits $i$ and different length values do not mask each other because each length turns a distinguished bit to set.
Encoding the length feature of the values introduces another benefit to the system:
Positioning the length segment as the left-most segment allows the system to use short circuit optimization, which skips unnecessary bit operations.
If there is no value in the candidate row with the same length as a query value, \system does not need to check the character segments.
%\ma{do we need the next example now? because we have the illustrating example now here?}
Consider our running example. The cell values ``Boxer'' and ``Birder'' in rows $5$ and $6$ respectively. Both of these values start with the encoded character ``B''. Therefore, ``B'' and its position cannot differentiate these two cell values. However, the values have different lengths, which makes their hashes distinguishable.

% Going back to our example, as $|C1|=8$, it will be encoded as $8\ mod\ 17 = 8$ and turn the eighth bit of the length segment.

\subsubsection{Bit rotation}\label{subsubsec:bitrotation}
As a final measure to differentiate hash values without using more $1$ bits than $i$, we reduce the likelihood of so-called random matches through rotation of the character segments.
A random match occurs when a key value that is partially masked by the individual hash functions of individual row values is masked by the aggregated super key.
For example, we could have a query key that contains rare characters that occur in two different columns. Or if one value shares the length of the key and the other only shares the characters of the key value that we are searching for.
To prevent these kinds of FPs, \hash rotates the character hash segments of a value $v$ by its length $l_v$ to left.
The most left bits that fall off will be moved to the beginning of the bit vector.
For example, a 3-bit rotation of '01100101' equals '00101011'. Note that the rotation only applies to the character-related segments.
%This rotation ensures that the hash of the same value is always rotated into the same bit structure.
As a result, a random match becomes less likely because the length of the random key value must match the length of the column that overlaps in terms of the least frequent characters.
%With bit rotation, \hash connects the value length and the character features without additional $1$ bits.
% Continuing our example in \ref{subsub_length}, we exploit the length of the values to re-locate the character bits so that the character ``B'' in ``Boxer'' is mapped to a completely different bit than the one for the same ``B'' in ``Birder''.

\textbf{Lemma.} Rotation reduces inter-column bit collisions. 
% \vspace{-.3cm}
\begin{proof}[Proof.]
Given that two cell values $c_1$ and $c_2$ are masked by a given super key $sk$, there are two general cases that lead to $sk$ masking false positives. In both cases the length bits must overlap. Either two values from the same domain or from different domains have the same length respectively. In all collision cases, the $K$ characters of each cell value $c_1$ and $c_2$ are masked, where $K = \alpha - 1$. 
Now assume that each of the $K$ characters is rotated based on the length of $c_1$ and $c_2$ respectively. Now, for a join candidate $\hat{c}_1$ and $\hat{c}_2$, the collision occurs only if there is an exact match of characters and length between any pair of $c_i$ and $\hat{c}_j$. In all other cases there is a 1-bit in a position that is not masked by $sk$.
We now prove that the opposite leads to a contradiction.
There is a collision under rotation of two cell values with different length where there is no exact match between any pair of $c_i$ and $\hat{c}_j$. It follows that $\forall c_i \exists \hat{c}_j: |c_i|=|\hat{c}_j|$ with an injective mapping. 
Thus, there is a pair of $c_i$ and $\hat{c}_j$ with the same length, where one of the encoded $K$ characters is different. Therefore $K-1$ character bits will be rotated the same way and are covered. Every other $c_i$ covers at most $K$ characters. Thus, there must be an exact match between $c_i$ and $\hat{c}_j$.
\end{proof}

The previous lemma shows that rotation reduces the number of collisions. This is also shown in our micro-benchmarks where we analyze its impact on filtering.
Continuing our illustrating example, the character segment bits all rotate by $8$ positions to the left.
As can be seen in the rotated array, the segment for ``u'' moved from the 108th to the 100th most left bit.
The final aggregated results for C2 and C3 are shown in separate hash array excerpts.
Note that because of the rotation in the hash results, the character ``u'' appears in different locations for $C_1$ and $C_2$ respectively.

\subsection{Index Updates}
There are three possible types of edits on table corpora that lead to updates in the index-level: {\em insert}, {\em update}, and {\em delete}.
Inserting a new table to the corpus requires the following updates.
Other than generating the PL items for the cell values in the newly added table, a super key is generated for each row. 
Inserting a new row to an existing table also follows the same procedure. 
Adding a new column to an existing table requires applying \hash on each individual column value and replacing the corresponding super key by the result of a bit-wise \texttt{OR} operation with the new \hash result.
Updating the value of a cell requires a complete re-hash of the corresponding super key.
Deleting a table/row only requires deleting the PL items for the table/row.
Yet, deleting a column from a table might change the super key entries, triggering a rehashing of all rows.
Although some of the aforementioned updates require regeneration of the super key, the system can handle the changes locally to the affected table.
%Generally, the most frequent index updates in a data lake would be \textit{table inserts}, which do not affect any of the already generated index values.
%!TEX root = ../main.tex

\section{Joinable Tables Discovery}\label{sec:index_applicaiton}
We now discuss how we leverage our index structure and \hash to efficiently find n-ary joins.
Recall the discovery phase is comprised of four main steps (see Figure~\ref{fig:architecture}): initialization, table filtering, row filtering, and joinability calculation.
Algorithm~\ref{alg:application} shows the pseudocode of this discovery process.
Overall, \system receives a query table $d$, a set of columns from the query table as the composite key $Q$, and an integer $k$ (Line~\ref{code:inputs}) with the goal of discovering the top-k joinable tables for $d$ and based on $Q$.

\begin{algorithm} [t!]
\footnotesize
\textbf{Inputs:}
$d$: query table, $Q$: set of query columns, $k$: the maximum number of desired tables \label{code:inputs}

TOPK $\leftarrow$ empty heap \label{code:empty_heap}

q\textquotesingle $\leftarrow$  init\_column\_selection(d, Q) \label{code:ics}

list\_of\_all\_PLs $\leftarrow$ fetch\_PLs(q\textquotesingle) \label{code:fetch}

candidate\_tables $\leftarrow$ extractTable\_sorted\_ids(list\_of\_all\_PLs) \label{code:table_grouping}

superkey\_map\_Q$\leftarrow$ map(Q, generate\_super\_keys(Q, d)) \label{code:queryhashing}

\For {$t$ in candidate\_tables} {\label{code:tableforloop}

table\_PLs $\leftarrow$ extractPLs($t$, list\_of\_all\_PLs) \label{code:get_table_pls}

\If {TOPK==k AND size(table\_PLs) < TOPK.min().$\jmath$}{ \label{code:tablefiltering1}

			BREAK and GOTO Line \ref{code:return};\label{code:tablefiltering1_goto}
}			

		$r_{\text{checked}}$ $\leftarrow$  0 \label{code:init_checked_vals}

	$r_{\text{match}}$ $\leftarrow$ [] \label{code:init_candidates}
	
	\For {pl\_item in table\_PLs}{\label{code:for_pl_items}

			\If {TOPK==k AND size(table\_PLs) - $r_{\text{checked}}$ + size($r_{\text{match}}$) < 
			TOPK.min().$\jmath$}{\label{code:tablefiltering2}
			
				Break and GOTO Line \label{code:tablefiltering2_goto}\ref{code:tableforloop};
			}\label{code:end_filtering2}
			d\_rows $\leftarrow$ superkey\_map\_Q.get(pl\_item.value) \label{code:corresponding_rows}
			
			\For {d\_row in d\_rows} {\label{code:for_corresponding_rows}
			
				\If {d\_row.superkey $\vee$ pl\_item.superkey = pl\_item.superkey} {\label{code:row_filtering}
			
					candidate\_rows.add(pl\_item)\label{code:add_into_candidate}
			}}
			checked\_values += 1 \label{code:increment_checked_vals}
	}
	$\jmath$ $\leftarrow$ calculateJ(candidate\_rows) \label{code:calc_jscore}
	
	TOPK.update(table\_id, $\jmath$)\label{code:topk_update}
}
return TOPK \label{code:return}

%\caption{\system algorithm (online phase).}
\caption{\system discovery algorithm.}
\label{alg:application}

\end{algorithm}

% \begin{algorithm} [t!]
% \textbf{Inputs:}
% \newline
% \textit{D}: input dataset\newline
% \textit{Q}: set of query columns\newline
% \textit{k}: the final number of joinable tables

% ic $\leftarrow$ init\_column\_selection(D, Q) \label{code:init_column_selection}

% candidate\_tables $\leftarrow$ Fetch\_pl(D, Q[ic]) \label{code:fetching}

% input\_super\_key\_generation() \label{code:preprocessing}

% \For {table in candidate\_tables}{\label{code:forall_tables}

% \If {table\_filtering\_1(top\_k, table)}{\label{code:table_filtering1}
        
%             break;\label{code:break1}
%         }

% \For {item in table.PL\_items}{\label{code:forall_items}
% \If {table\_filtering\_2(top\_k, table, candidates)}{\label{code:table_filtering2}
        
%             break;\label{code:break2}
%         }
% input\_rows $\leftarrow$ corresponding\_input\_super\_keys(item)\label{code:corresponding_rows}

% \For{input\_row in input\_rows}{\label{code:forall_rows}

% \If {input\_row.superkey $\lor$ item.superkey == item.superkey}{\label{code:mate_filtering}
        
%             candidates.Add(input\_row, item)\label{code:add_candidate}
%         }
% }
% }
% score $\leftarrow$ joinability(candidates) \label{code:joinability}

% Update(score, top\_k)\label{code:update_topk}
% }

% return top\_k
% %\caption{\system algorithm (online phase).}
% \caption{\system discovery algorithm.}
% \label{alg:application}

% \end{algorithm}

In the {\em initialization} step, \system selects the initial column, fetches the candidate tables, and applies our hash procedure, i.e.,~super key generation, on the join columns of the input table (Lines~\ref{code:fetch}-\ref{code:queryhashing}).
In the {\em table filtering} step, it leverages the number of corresponding index hits per table to prune the tables that cannot be among the top joinable tables (Lines~\ref{code:tablefiltering1}-\ref{code:end_filtering2}).
We introduce two coarse-grained pruning rules to filter non-promising tables while retaining all relevant tables.
In the {\em row filtering} step, the system prunes as many non-joinable rows as possible for each candidate table (Lines~\ref{code:corresponding_rows}-\ref{code:increment_checked_vals}).
It evaluates the rows of candidate tables that are likely to be among the top-k tables, by only checking the membership of the key values through the generated super keys.
We detail the aforementioned three steps in the following subsections.

In the final {\em calculateJ} step, \system validates the rows that remain after the previous two filtering steps and calculates the joinability score ($\jmath$) of the remaining tables as defined in Section~\ref{sec:problem_statement}.
To do so, it fetches the actual cell values of each row, compares them to the composite key values (Line~\ref{code:calc_jscore}).
It then updates the list of the top-$k$ joinable tables (Line~\ref{code:topk_update}).
\system terminates and reports the discovered joinable tables (Line~\ref{code:return}), once it has discovered the top-$k$ joinable tables or checked all the candidate tables.

\subsection{Initialization Step}
\system has to initiate the retrieval with one of the key columns in $Q$ using the single column index.
The initial column can heavily impact the runtime of the system.
Consider the Kaggle IMDB dataset\footnote{\url{https://www.kaggle.com/carolzhangdc/imdb-5000-movie-dataset}}, which contains information about more than $5000$ movies. If we use it as the query table with the join key <``Director name'', ``Movie title''>, we obtain 500k tables when fetching with the column ``Director, and 38M tables when using ``Movie Title'', respectively.

Yet, selecting the ideal initial query column depends on how many PL items each cell value will match and cannot be known upfront.
We, thus, resort to a heuristic that only considers the query table at hand.
We choose the column with the smallest number of unique values, i.e.,~lower cardinality, hoping to match fewer PLs in the index.
%Our IMDB example confirms our intuition as column ``Movie Title'' contains \textit{5043} and ``Director Name'' contains \textit{2400} unique values.
Thus, \texttt{init\_column\_selection()} picks the column with the minimum cardinality, $q\textquotesingle \in Q$ (Line~\ref{code:ics}). 
With the initial column, \system fetches PL items including their generated super key for the corresponding rows (Line~\ref{code:fetch}).
Then, the fetched PL items are grouped based on the table identifiers~(Line~\ref{code:table_grouping}).
At the same time, it sorts the tables based on the number of PL items in each table to evaluate more promising tables first and enabling table pruning techniques.
Next, \system generates the relevant super keys for the join columns of the query table.
For each column $q \in Q$, it hashes the column values and uses an \texttt{OR} operation to generate the aggregated hash  for each key value combination (Line~\ref{code:queryhashing}).
We generate a dictionary to map initial query column values to the generated super keys.
This helps us to efficiently align fetched super keys with query table super keys in later steps.

\subsection{Table Filtering}
%Although selecting the optimal query column can reduce the number of candidate tables, there are still many tables that need to be verified for top-k selection.
We propose a {\em coarse-grained} table filtering approach, inspired by prefix-filtering~\cite{chaudhuri2006primitive}, to further reduce the set of candidate tables.
While traversing over the fetched tables, \system prunes tables that do not stand a chance to be selected as the top-k joinable candidates.
For each candidate table (Line~\ref{code:tableforloop}), it leverages the corresponding PL items of the table (Line~\ref{code:get_table_pls}).
Table filtering applies two strategies:

\noindent\textit{(1) } If $\jmath_k$ is the joinability score of the worst table in the top-k list, i.e.,~table with the lowest joinability score, and $L_t$ is the number of PL items for current candidate table $t$, the system drops the candidate table $t$ from further evaluations iff $L_t \leq \jmath_k$ (Line~\ref{code:tablefiltering1}).
As tables are sorted in decreasing order based on their number of PL items, the remaining tables contain equal or smaller number of PL items and cannot end up among the top-k tables. Thus, the discovery task halts when this rule activates for one table (Line~\ref{code:tablefiltering1_goto}). 
Going back to our running example, assume that we aim to discover the top one joinable table to $d$ and we find $T_1$ as a joinable table with joinability of $\jmath_k = 5$. The moment that a candidate table arrives for evaluation and it only contains $4$ or less PL items, the system can return $T_1$ as the most joinable table. This is because the new and the following candidate tables in the best case, where all the rows are joinable, will have the joinability of $4$ and cannot replace $T_1$.

\noindent\textit{(2) } If $r_{\text{checked}}$ is the number of already evaluated rows from a table $t$ and $r_{\text{match}}$ is the number of joinable rows among the evaluated rows, \system drops the current table iff $L_t - r_{\text{checked}} + r_{\text{match}} \leq \jmath_k$ (Line~\ref{code:tablefiltering2}).
This rule skips the current candidate table, because even if all the remaining rows are joinable to the input table, it cannot achieve a higher joinability score than the worst top-k table.
Already during this check, the system leverages the super key to filter joinable rows. Once a table is skipped, the algorithm evaluates the next table (Line~\ref{code:tablefiltering2_goto}).
Both table filtering strategies only apply after \system already saw $k$ joinable tables.
Once again, assume the previous example, where the user wants to find the top one joinable table to $d$ and $T_1$ with $\jmath_k = 5$ is already discovered. Previous rule cannot drop a candidate table with $10$ PL items. Now, assume that \system evaluates the first $7$ PL items of this newcomer table (See Row Filtering in~\ref{subsec:rowfiltering}) and only one of them are in fact joinable to $d$. In this case, the system can ignore the remaining PL items because even if the remaining three rows are joinable, the candidate table only reaches the joinability of $4$ that does not replace $T_1$.

\subsection{Row Filtering}\label{subsec:rowfiltering}
If there is a prospect of a candidate table to be one of the top-k tables, meaning that the candidate table incorporates enough candidate joinable rows, \system evaluates them. 
It leverages the super keys generated for both rows in the candidate and query tables to efficiently check the membership of the key values in the candidate rows.
If the super key of a query table row is not masked by the super key from the candidate joinable table, the system can simply drop the table row from joinability computations. 

For each candidate row, represented by the PL item (Line~\ref{code:for_pl_items}), \system discovers which query rows should be compared with the current candidate row.
To find the query rows efficiently, the system uses the generated dictionary that maps the initial column values to the corresponding super keys of the composite key value combination (Line~\ref{code:corresponding_rows}). In our running example, row $1$ in $T_1$ is a candidate row and \system binds it to the $5^{th}$ row in $d$ using a dictionary that maps the cell values to the input rows. This step ensures that we do not iterate over all query table rows for each candidate row. Then, for each corresponding input row \texttt{d\_row} (Line~\ref{code:for_corresponding_rows}), the super keys of the current candidate row and its corresponding input rows are compared through bit-wise \texttt{OR} masking (Line~\ref{code:row_filtering}).
An input row is a join candidate to its corresponding row of the PL item iff the result of the bit-wise \texttt{OR} operation on their super keys is equal with the super key of the PL item (Line~\ref{code:add_into_candidate}).
%This is because the values in the composite key should be a subset of the values in the joinable row in the joinable table.
\newline
\noindent\textit{Example 3.} 
Consider our running example once again. If the hash values for ``Muhammad'', ``Lee'', ``US'', ``Ali'', ``Germany'', ``Dancer'', ``Boxer'', and ``Birder'' are $1001000$, $01100000$, $00010100$, $00010001$, 
$10001001$, $10000010$, $10000001$, and $00001001$ respectively, the aggregated super key for the rows with value ``Muhammad'', i.e., rows $2$, $5$, and $6$ in the Table~$T_1$, would be $11111110$, $11011101$ and $11101001$ respectively.
Likewise the super keys over the columns in $Q$ and for the first row in \textit{d} would be $1001000 \vee 01100000 \vee 00010100 = 1111100$.
In our example, the resulting super key $1111100$ is not a subset of $11011101$ and $11101001$, which are the super keys of rows $5$ and $6$. Therefore, $5^{th}$ and $6^{th}$ rows are removed from further $\jmath$ calculation.
On the other hand, $1111100$ is subsumed by $11111110$, i.e., the super key of the second row of $T_1$.
Therefore, we add the second row to the list of candidate rows.
Similar to bloom filters, this operation can lead to FPs but never to false negatives (FNs). 
%After this comparison, \texttt{checked\_values} is increased for the next table filtering check.

\textbf{Lemma.} The super key will never miss a joinable row.
    % \vspace{-.3cm}
\begin{proof}[Proof by contradiction.]
Given a table $R$ with a row $r$ where $c_1$, $c_2$,... are the cell values of the row $r$. Now assume that we have a deterministic hash function $h$ so that $h(c_1) =h_1, h(c_2) = h_2, \dots$. The super key $sk$ is constructed as follows: $sk = h_1 \lor h_2 \lor … \lor h_n.$
Now assume, there is a table $T$ with a 2-column join key and a row with a value combination ${k1, k2}$ with $k1=c1$ and $k2=c2$, so that the columns are joinable with $r$ but the bitwise OR operation of $h(k_1) \lor h(h_2)$ will not be covered by $sk$, i.e.,
$h(k_1) \lor h(k_2) \lor sk \neq sk$

However by construction $h_1 \lor h_1 \lor sk = sk$.
It follows $h_1 \neq h(k_1) \lor h_2 \neq h(k_2)$, which is a contradiction to $k1=c1$ and $k2=c2$. 
\end{proof}

\subsection{Analysis} \label{subsec:proof}
There are two main features of \hash that lead to higher efficiency over BF:  1) The non-uniform distribution of hash values; 2) the encoding of the length. First, unlike BF that is built on the premise of uniform distribution of its underlying hash functions, 
\hash exploits the knowledge that different values come from different columns. Building on the premise that each domain has unique syntactic features~\cite{DBLP:journals/pvldb/ZhangSLHDT20}, \hash maps hash results to unique segments based on their domain to prevent random overlaps.
Assume an $|a|$-bit hash array. The probability of a collision of two words using LHBF~\cite{DBLP:conf/esa/KirschM06} is $\frac{1}{\binom{|a|}{2}}= \frac{2}{|a|\cdot (|a|-1)}$.
Now consider \hash. To generate a collision one has to hash a value with the same set of $K$ rare characters and the same length, where $K = \alpha -1$. The probability of drawing the same $K$ characters out of 37 in the same set of $\beta$ positions is $\frac{1}{37\cdot\beta}\cdot \frac{1}{36\cdot\beta} \dots \cdot \frac{1}{(37-K+1)\cdot \beta}$. We now solve the inequality $\frac{2}{|a|\cdot (|a|-1)}>\frac{1}{\beta}\cdot \prod_{i=1}^{K}{\frac{1}{37-K+1}}$. For  $|a|=128$ ($ \beta = 3$), the inequality is true for $K>3$ . If we now also include the information about length encoded in the remaining $17$ bits we obtain $\frac{1}{|a|\cdot (|a|-1)}>\frac{1}{17}\cdot \frac{1}{3}\cdot \prod_{i=1}^{K}{\frac{1}{37-K+1}}$, which is already true for $K>2$. Thus, our approach that explicitly leverages character positions and the length of values leads to fewer collisions when encoding $K$ random characters. Note that the rarity of the characters was not included in this setup. 

%!TEX root = ../main.tex
\section{Experiments}\label{sec:experiments}
We evaluate \system having the following questions in mind:
(i)~{\em What is the runtime benefit of using \system compared to the state-of-the-art inverted index~\cite{abedjan2015dataxformer} when dealing with n-ary joins?}
(ii)~{\em How do parameters, such as the cardinality of tables, the choice of $k$ , $|Q|$, and the hash size affect the runtime?}
(iii)~{\em What is the filtering power of \hash compared to other hash functions?}
\subsection{Experimental Setup} \label{subsec:experimental_setup}
We probed different query tables against two different table corpora: the openly available Dresden Web Table Corpus (DWTC) and the German Open Data repository(\url{https://www.govdata.de/}). These corpora are selected because they yield tables with different sizes and dimensionalities.
The DWTC corpus contains more than $145M$ tables, $870M$ columns, $1.45B$ rows, and $660M$ unique values. 
The German open data also contains $17k$ tables, $440k$ columns, $62M$ rows, and $20M$ unique values. 
As query tables, we leverage sets of random inputs.
Similar to the state-of-the-art~\cite{zhu2019josie}, we selected $900$ query tables as six groups ($150$ tables in each category) from open data and webtable corpora. Each group is randomly selected from different cardinality distributions and we make sure that no duplicate tables are included inside those groups.The query columns are selected randomly.
For open data, the groups reflect query tables with cardinality smaller than $100$, $1000$, and $10k$, respectively. As WTDC tends to have smaller cardinalities, the groups contain tables where the cardinalities are smaller than $10$, $100$, and $1000$. In the experiments, we label the input query table collections as \textit{Corpus (size)}, e.g., \textit{WT (10)} and \textit{OD (100)}.
% The query sets on average, have the cardinality of $33$, $394$, $4933$, $3$, $17$, and $169$ and lead to joinability score $40$, $1434$, $8187$, $4$, $52$, and $99$ for OD (100), OD (1k), OD (10k), WT (10), WT (100), and WT (1k), respectively.
In addition, we also leverage \textit{School} corpus, generated in a previous work~\cite{chepurko13arda}. This corpus contains $335$ tables. The tables are comprised of over $27$ and $30,000$ columns and rows on average respectively. This corpus enables us to evaluate the systems in the existence of larger tables. In this set of experiments, we extend the tables containing ``Program Type'' and ``School Name'' columns with joinable tables.
We also selected machine learning datasets that require enrichment from Kaggle. This query set has more general content, therefore, they run against the webtable corpus. These query tables lead to almost $7M$ joinable candidates on average based on their first query column. Table~\ref{tab:queries} shows the query table sets used in this paper including statistics about the number of tables, average cardinality, and the average joinability score.
\begin{table}[]
    \scriptsize
    \centering
    \caption{Input query tables.}
    % \vspace{-2em}
    \label{tab:queries}
\begin{tabular}{l|l|l|l|l}
%\toprule
\textit{\textbf{Query Set}} & \textit{\textbf{\# of tables}} & \textit{\textbf{Corpus}} & \textit{\textbf{Cardinality}} & \textit{\textbf{Joinability}} \\ \toprule

WT (10) & 150 & DWTC & 3 & 4 \\
WT (100) & 150 & DWTC & 16 & 52 \\
WT (1000) & 150 & DWTC & 151 & 99 \\
OD (100) & 150 & German Open Data & 15 & 40 \\
OD (1000) & 150 & German Open Data & 263 & 1,434 \\
OD (10000) & 150 & German Open Data & 2455 & 8,187 \\
Kaggle & 11 & DWTC & 34,400 & 2,318 \\
School & 10  & School corpus & 3,100 & 15,130 \\

\end{tabular}
\end{table}

In all our experiments, unless specified otherwise, we leverage the $128$-bit hash space and we look for top-$10$ joinable tables.
We implemented \system in \textit{Python 3.7} and ran it on a server with 128 CPU cores, 500 GBs of RAM, and 2 TBs of SSD drive. 
We used Vertica \textit{v10.1.1-0}~\cite{lamb2012vertica} to store the index.
Our code and datasets are publicly available~\footnote{\url{https://github.com/LUH-DBS/MATE}}.

\noindent{\textbf{Index generation.}} 
All of the hash-based solutions including \system+\hash require a prior hash generation. 
The original webtable and open data corpus sizes are approximately $250$ and $12$ GB each with the inverted index SCR.
\system requires additional space to store the super keys. As we defined the super keys per cell value (See Section~\ref{sec:desiderata}), \system needs $8.3B \times 128b = 123.6$ and $62M \times 128b = 11.9$ Gigabyte for webtables and open data, respectively.
Leveraging more efficient structure, where the super keys are stored per row, one can reduce the index size to $1.45B \times 128b = 21.6$ and $62M \times 128b = 0.92$ Gigabyte, respectively. This space efficient structure might lead to integration overhead, where the super keys are joined with the PLs.
% As super keys are defined per row, using $128$ bits hash space, \system needs $1.45B \times 128b = 21.6$ and $62M \times 128b = 0.92$ Gigabyte for webtables and open data, respectively. 
%For $256$- and $512$-bit hash sizes the additional required space doubles and quadruples, respectively.
% To avoid the overhead of finding the corresponding row per value, we store each super key for each row value. This increases the index size for webtables and open data to GB and GB, respectively. 
Josie on the other hand, requires $293$ and $20$ Gigabytes additional storage for webtables and open data, respectively. Note that the index for Josie is not sufficient for multi-column join discovery as it does not store the rows information of the cell values. Therefore, Josie-based approaches require an inverted index like SCR. 
The offline index generation for \system takes $35$ and $2$ hours for webtable and open data corpora, respectively. For Josie, the index generation time is $336$ and $50$ hours, respectively.

\subsubsection{Systems}\label{subsec:baselines_systems}
To our best knowledge, there is no a multi-attribute join discovery system. We compare \system and \hash, to multiple possible adaptations of single-column systems. 

\noindent{\textbf{Single-Column Retrieval (SCR).}} 
We, adapted a single column index SCR for n-ary join discovery. It uses all the optimizations in Algorithm~\ref{alg:application} including the table filtering strategies and initial column selection.
Yet, SCR cannot utilize the super key:
It fetches the candidate joinable rows and validates them through exact value comparisons in memory. %We have tested two versions of SCR: when the data fits into memory as has presented in~\cite{zhu2019josie} and the data does not fit into memory following the implementation of DataXFormer~\cite{abedjan2015dataxformer}. The two versions only differ in the initial fetching time for the tables that are joinable in the initial column.

\noindent{\textbf{SCR Josie.} Josie is the a state-of-the-art single-column join discovery algorithm introduced in literature. Josie leverages an index that maps values to columns, where as the columns are represented as sets. To infer the joinable rows we fall back on the SCR index.}

\noindent{\textbf{Multi-Column Retrieval (MCR)~\cite{zhu2019josie}.} In this approach, we fetch the PL items for each query column and intersect the results to discover the top-k joinable tables.}

\noindent{\textbf{MCR Josie~\cite{zhu2019josie}.} One can adapt Josie to discover multi-column joins by repeating it on every key column in Q, finding the most joinable tables per column, and then intersecting them and evaluating the tables that appear in all joinable results. 
% Due to the fact that there is no guarantee that the top-k joinable tables based on the composite key appears in the top-k of all joinable tables for single-columns this approach can lead to false negatives. In the simplest scenario $k = 1$. The goal is to find top-1 joinable table based on ``F. Name'', ``L. Name'', and ``Country''. One cannot make sure that this top-1 joinable table is the same with the top-1 joinable table based on ``Country''. In this baseline, to prevent FNs, we set $k = 500$ for the initial query column and a large $k = 10k$ for the remaining columns. 2) The offline index generation is very time consuming specially for large corpora such as, DWTC.
% 2) Repeating Josie per column is more time consuming specially if the number of key columns is large. 
}

\subsubsection{Hash Functions} \label{subsec_baseline_hash}
We compare the filtering performance of \system using different hash functions.

\noindent{\textbf{Bloom filter (BF).}} This is a variation of \system that uses bloom filters (instead of \hash) with a fixed number of hash functions to generate the super key~\footnote{\url{https://gist.github.com/malaikannan/22401e67b332f52651dcef8ea5039498}}.
The BF implementation calculates the number of hash functions based on the average number of columns in the corpus tables. We use Murmur3 hash family as the base function in the BF implementation. 
In BF, the FP ratio can be calculate as: $FP = (1-e^{-\frac{V\cdot H}{|a|}})^H$, where $V$ is the number of values to be inserted and $H$ is the number of hash functions~\cite{fan2000summary}. To calculate the optimum number of hash functions, we set FP to $0$. Doing so, the number of hashes can be calculated as: $H = \frac{|a|}{V} log2$.
We set $V=5$ for webtables and $V=26$ for OD, using the average number of columns in the corresponding corpus.

\noindent{\textbf{Less Hashing Bloom Filter~\cite{DBLP:conf/esa/KirschM06} (LHBF).} This approach is an optimized version of the standard bloom filter that uses only two hash functions per value. We use the LessHash-BloomFilter python library, which is publicly available.}~\footnote{\url{https://github.com/garawalid/LH-BloomFilter}}

\noindent{\textbf{Hash table (HT).}} 
It is the same as BF but uses one hash function. 

\noindent{\textbf{Other hash functions.}} 
We also compare \system to hash functions that are not followed by any post-processing of the hash values. 
We considered the following state-of-the-art hash functions: SimHash~\cite{charikar2002similarity, sadowski2007simhash}, MD5~\cite{rivest1992md5}, Google's CityHash~\cite{pike2017cityhash}, and Murmur.
%These functions are fast and lead to low collision rates. %Therefore, they are suitable for membership lookup tasks.
We excluded hash functions based on SHA as they turned out to be similar but slower than the aforementioned hash functions.

\subsection{\system VS. Baseline Systems}
\begin{figure}[t!]
\centering
\begin{tikzpicture}
  \begin{groupplot}[xtick=data, ybar, enlarge x limits=0.1, symbolic x coords={WT (10), WT (100), WT (1000), OD (100), OD (1000), OD (10000)}, ymin = 0, group style={group size=1 by 1 , horizontal sep=.0cm}, height=4.8cm, width=8cm, ymax = 300, every node near coord/.append style={yshift=-0.21cm}, point meta=explicit symbolic, log origin=infty, xticklabel style={font=\footnotesize, rotate=25}, xmajorgrids=true, ymode = log, ytick={1, 10, 100, 1000, 10000},
    	ymajorgrids=true,grid style=dashed ]
    \nextgroupplot [ylabel={Runtime (seconds)}, 
    legend style={legend columns=5,at={(0.5,-0.3)},anchor=north,font=\footnotesize}
    , bar width=3pt]

    \addplot[blue,fill] coordinates {(WT (10), 1.8) (WT (100), 2.0) (WT (1000), 1.6) (OD (100),3.1) (OD (1000), 4.8) (OD (10000), 13.6)};\addlegendentry{\hash (128 bits)};
    
     \addplot[purple,fill] coordinates {(WT (10), 72.4) (WT (100), 51) (WT (1000), 97.6) (OD (100), 6.3) (OD (1000), 4.3) (OD (10000), 14.40)}; \addlegendentry{MCR};
    
    \addplot[red,fill] coordinates {(WT (10), 15.5) (WT (100), 25.4) (WT (1000), 33.5) (OD (100), 22.9) (OD (1000), 52.0) (OD (10000), 175.6)}; \addlegendentry{SCR};
    
    \addplot[brown,fill] coordinates {(WT (10), 300) (WT (100), 300) (WT (1000), 300)(OD (100), 9.4) (OD (1000), 42.0) (OD (10000), 71)}; \addlegendentry{MCR Josie};
    
    \addplot[gray,fill] coordinates {(WT (10), 16.6) (WT (100), 27.3) (WT (1000), 35) (OD (100), 24.7) (OD (1000), 60) (OD (10000), 191.8)}; \addlegendentry{SCR Josie};
    
  \end{groupplot}
\end{tikzpicture}
% \vspace{-.4cm}
\caption{Runtime comparison between \system and SCI.}
\label{fig:dxf_vs_mate_runtime}
%\vspace{-.2cm}
\end{figure}
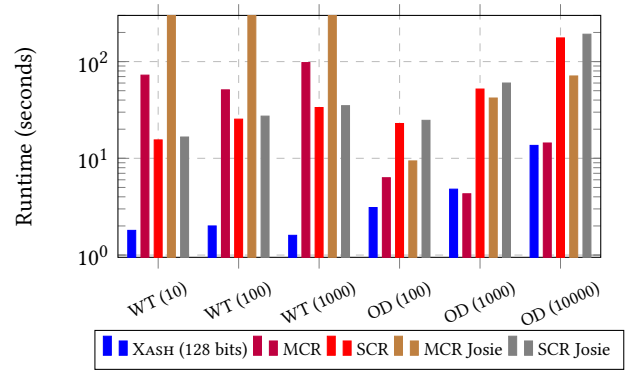
%Here, we only report the runtime for the 512-bit version of \system.
%Later in this section, we also evaluate the system performance for hash arrays with varied sizes.
Figure~\ref{fig:dxf_vs_mate_runtime} shows the runtime comparison \system against SCR, MCR, and the corresponding JOSIE implementations in log scale.
Without the row filtering optimization of \system, the runtime for joinability calculation dominates the overall runtime.
The depicted runtime results for baselines include the time required to discover the multi-column joins in-memory on top of the state-of-the-art single-attribute joinable table discovery systems. 
Note that there is an additional fetching cost that is negligible, i.e.,~in the order of milliseconds when we use the in-memory implementation like state-of-the-art~\cite{zhu2019josie}. However, the fetching cost can vary between $1$ and $40$ seconds when the data and the index has to be retrieved from disk. This is the case for the DWTC, which cannot fit into memory. The depicted runtime does not contain the fetching time as it is the same for both approaches.
We observe that \system outperforms baseline systems in almost all experiments: it is up to $61x$, $13x$, $9x$, $22x$ faster than MCR, SCR, MCR Josie, and SCR Josie respectively.
%As a result, it can drop most of the non-joinable table rows before any further computation.
This experiment also shows that no other baseline constantly performs better than the other approaches. For instance, SCR-based approaches are slower than their corresponding MCR-based systems for open data queries but on the large webtable corpus they underperform. Note that MCR Josie for webtables did not terminate in $7$ days. The slower performance of the MCR approaches on webtable queries is because accessing the relevant PLs does not scale for the size of the webtable corpus. While MCR Josie performs similar to MCR in OD (100) its performance drops as the cardinality of the open data queries grows.
The size of the input dataset correlates with the runtime. This is expected because join columns with higher cardinality initially match more PL items in the index which again need to be fetched and filtered. 

In addition to the input table size, FPs play an important role in the number of comparisons and ultimately the runtime.
For instance, in case of SCR, although OD (1k) and WT (1k) have similar query table sizes, the OD (1k) queries lead to $35\%$ higher runtime than WT (1k). This increase in runtime is because of the large number of FPs for OD (1k) tables, i.e., $3M$ more FP rows compared to WT (1k).
%Moreover, For some query datasets, such as OD (10,000), the performance difference between \system and baselines is smaller, because the initial column filters already many of the FPs.
%On the other hand, for some datasets, such as WT (10), this performance gap is substantial because SCR retrieves over one orders of magnitude more FPs than \system. % 400 times more
\newline
\textbf{Summary.}
(i) \system is up to $100x$ faster than unary join discovery systems in discovering n-ary joins.
(ii) Performance gain of \system over SCR-based approaches depends on the number of FP rows.

\subsection{\hash VS. Baseline Hash Functions} \label{subsubsec:xash_vs_hashes}
Table~\ref{tab:dataset_runtime} depicts the runtime of \system with different hash functions, including the proposed \hash.
Note that all the competing hash functions benefit from all of \system's optimizations and only differ in the applied hash function during row filtering.

\begin{table*}[]
     \footnotesize
    \centering
    \caption{Runtime experiment (seconds). \colorbox[HTML]{A2EDFF}{blue} cells represent the experiments where the larger hash performs worse. \colorbox[HTML]{FFABA8}{Red} cells show the maximum performance gain of \system over BF.}
    % \vspace{-.2cm}
    \label{tab:dataset_runtime}
\begin{tabular}{r|r|r|r|r|rrr|rrr|rrr|rrr|rrr}
%\toprule
\textit{\textbf{Dataset}} &
\textit{\textbf{SCR}}&\textit{\textbf{\thead{MD5}}}&\textit{\textbf{{Murmur}}}&\textit{\textbf{{City}}}&
 \multicolumn{3}{c|}{\textit{\textbf{{SimHash}}}}& \multicolumn{3}{c|}{\textit{\textbf{{HT}}}}& \multicolumn{3}{c|}{\textit{\textbf{{BF}}}}& \multicolumn{3}{c|}{\textit{\textbf{{LHBF}}}}&  \multicolumn{3}{c}{\textit{\textbf{{\hash}}}}
 \\

&&\multicolumn{1}{c|}{\textbf{128}}&\multicolumn{1}{c|}{\textbf{128}}&\multicolumn{1}{c|}{\textbf{128}}&\multicolumn{1}{c}{\textbf{128}}&\multicolumn{1}{c}{\textbf{256}}&\multicolumn{1}{c|}{\textbf{512}}&\multicolumn{1}{c}{\textbf{128}}&\multicolumn{1}{c}{\textbf{256}}&\multicolumn{1}{c|}{\textbf{512}}&\multicolumn{1}{c}{\textbf{128}}&\multicolumn{1}{c}{\textbf{256}}&\multicolumn{1}{c|}{\textbf{512}}&\multicolumn{1}{c}{\textbf{128}}&\multicolumn{1}{c}{\textbf{256}}&\multicolumn{1}{c|}{\textbf{512}}&\multicolumn{1}{c}{\textbf{128}}&\multicolumn{1}{c}{\textbf{256}}&\multicolumn{1}{c}{\textbf{512}}\\
\toprule
WT (10) & 20.6 & 12.5 & 12.6 & 12.6 & 11.1 & 10.1 & \colorbox[HTML]{A2EDFF}{12.5} & 6.4 & \colorbox[HTML]{A2EDFF}{7.3} & \colorbox[HTML]{A2EDFF}{7.6} & 4.2 & 4.2 & \colorbox[HTML]{A2EDFF}{10.3} & 4.1 & 3.6 & 2.2 & \textbf{1.8} & \textbf{1.5} & \textbf{1.4} \\
WT (100) & 24.5 & 15.3 & 14.8 & 15.3 & 13.4 & 8.5 & \colorbox[HTML]{A2EDFF}{17.2} & 9.5 & \colorbox[HTML]{A2EDFF}{10.1} & \colorbox[HTML]{A2EDFF}{12.4} & 4.7 & \colorbox[HTML]{A2EDFF}{5.0} & \colorbox[HTML]{A2EDFF}{16.8} & 5.6 & 3.6 & \colorbox[HTML]{A2EDFF}{4.2} & \textbf{1.6} & \colorbox[HTML]{A2EDFF}{\textbf{1.8}} & \colorbox[HTML]{A2EDFF}{\textbf{2.0}}\\
WT (1k) & 32.7 & 20.2 & 19.8 & 19.8 & 16.3 & 10.0 & \colorbox[HTML]{A2EDFF}{17.0} & 13.4 & \colorbox[HTML]{A2EDFF}{13.8} & 13.4 & 5.6 & 5.5 & \colorbox[HTML]{FFABA8}{16.7} & 7.6 & 4.1 & \colorbox[HTML]{A2EDFF}{4.6} & \textbf{1.6} & \textbf{1.5} & \colorbox[HTML]{FFABA8}{\textbf{1.6}}\\
\toprule
OD (100) & 247.6 & 212.1 & 224.2 & 194.5 & 102.9 & 69.1 & 56.5 & 24.0 & 16.4 & 10.8 & 11.0 & 4.8 & 3.4 & 30.1 & 9.5 & 8.3 & \textbf{6.5} & \textbf{3.5} & \textbf{3.}1\\
OD (1k) & 165.8 & 146.8 & 152.3 & 138.6 & 90.0 & 73.6 & 62.6 & 25.4 & 21.6 & 16.5 & 14.4 & 7.6 & 5.6 & 39.2 & 11.3 & \colorbox[HTML]{A2EDFF}{13.0} & \textbf{8.7} & \textbf{4.5} & \colorbox[HTML]{A2EDFF}{\textbf{4.8}}\\
OD (10k) & 256.7 & 203.9 & 202.1 & 190.6 & 145.6 & 127.0 & 103.5 & 49.6 & 44.0 & 31.9 & 27.7 & 17.8 & 14.3 & 79.2 & 27.0 & 21.8 & \textbf{17.6} & \textbf{13.4} & \colorbox[HTML]{A2EDFF}{\textbf{13.6}}\\
\toprule
Kaggle & 297.0 & 97.6 &104.2&104.4&{76.6}&57.3&41.0&34.8&30.3&25.9&19.7&14.8&13.1 & 37.8&27.1&21.8&\textbf{15.2}&\textbf{12.0}&{\textbf{12.3}}\\
\toprule
School & 873.6 & 772.7 & 772.9 & 742.8 & 593.7&444.0&307.2 & 54.3&38.3&31.0 & 24.2&19.0& 19.0 & 33.8&25.3&22.7 &\textbf{20.0}&\textbf{17.9}&\textbf{17.6}\\
\end{tabular}
\end{table*}

\begin{table*}[]
     \footnotesize
    \centering
    \caption{Precision experiment.}
    % \vspace{-.2cm}
    \label{tab:dataset_precision}
\begin{tabular}{r|c|c|rr|rr|rr|rr|rr}
%\toprule
\textit{\textbf{Dataset}} &
\textit{\textbf{\thead{MD5}}}&\textit{\textbf{{CityHash}}}&
 \multicolumn{2}{c|}{\textit{\textbf{{SimHash}}}}& \multicolumn{2}{c|}{\textit{\textbf{{HT}}}}& \multicolumn{2}{c|}{\textit{\textbf{{BF}}}}& \multicolumn{2}{c|}{\textit{\textbf{{LHBF}}}}&  \multicolumn{2}{c}{\textit{\textbf{{\hash}}}}
 \\

&\multicolumn{1}{c|}{\textbf{128}}&\multicolumn{1}{c|}{\textbf{128}}&\multicolumn{1}{c}{\textbf{128}}&\multicolumn{1}{c|}{\textbf{512}}&\multicolumn{1}{c}{\textbf{128}}&\multicolumn{1}{c|}{\textbf{512}}&\multicolumn{1}{c}{\textbf{128}}&\multicolumn{1}{c|}{\textbf{512}}&\multicolumn{1}{c}{\textbf{128}}&\multicolumn{1}{c|}{\textbf{512}}&\multicolumn{1}{c}{\textbf{128}}&\multicolumn{1}{c}{\textbf{512}}\\
\toprule
WT (10) & 0.27$\pm$0.39 & 0.25$\pm$0.38 & 0.28$\pm$0.40 &0.31$\pm$0.42& 0.34$\pm$0.43 &0.28$\pm$0.43 & 0.44$\pm$0.46 &0.40$\pm$0.43 & 0.44$\pm$0.45 & 0.61$\pm$0.46& \textbf{0.57$\pm$0.46} & \textbf{0.88$\pm$0.26}   \\
WT (100) & 0.27$\pm$0.40 & 0.27$\pm$0.40 & 0.27$\pm$0.40 &0.27$\pm$0.39 & 0.34$\pm$0.41 &0.38$\pm$0.41 & 0.46$\pm$0.44 &0.34$\pm$0.41  & 0.45$\pm$0.43 & 0.63$\pm$0.44& \textbf{0.61$\pm$0.43} &\textbf{0.93$\pm$0.22} \\
WT (1k) & 0.24$\pm$0.36 & 0.24$\pm$0.36 & 0.25$\pm$0.37 & 0.28$\pm$0.35 &  0.40$\pm$0.37 &0.42$\pm$0.36 &  0.59$\pm$0.39 & 0.32$\pm$0.35&  0.52$\pm$0.38  &0.78$\pm$0.33 & \textbf{0.77$\pm$0.34} &\textbf{0.98$\pm$0.10}  \\
\toprule
OD (100) & 0.27$\pm$0.38 & 0.28$\pm$0.39 & 0.28$\pm$0.38 &0.32$\pm$0.41 & 0.43$\pm$0.40 &0.55$\pm$0.41 & \textbf{0.56$\pm$0.41} & 0.79$\pm$0.34& 0.45$\pm$0.43 &0.67$\pm$0.35 & 0.52$\pm$0.41 & \textbf{0.80$\pm$0.34}   \\
OD (1k) & 0.32$\pm$0.40 & 0.32$\pm$0.40 & 0.32$\pm$0.39 &0.41$\pm$0.41 & 0.47$\pm$0.40 &0.59$\pm$0.41 & \textbf{0.61$\pm$0.40} &0.85$\pm$0.28 & 0.44$\pm$0.34 & 0.63$\pm$0.39 & 0.53$\pm$0.41 & \textbf{0.86$\pm$0.28}   \\
OD (10k) & 0.27$\pm$0.38 & 0.28$\pm$0.38 & 0.28$\pm$0.37 & 0.34$\pm$0.39& 0.42$\pm$0.39 &0.56$\pm$0.39 & \textbf{0.59$\pm$0.40} & \textbf{0.87$\pm$0.27}& 0.40$\pm$0.42 & 0.66$\pm$0.40& 0.52$\pm$0.42 & 0.82$\pm$0.32   \\
\toprule
School & 0.00$\pm$0.00&   0.00$\pm$0.00&   0.00$\pm$0.00 & 0.00$\pm$0.00 & 0.01$\pm$0.01& 0.03$\pm$0.02 & 0.07$\pm$0.03&  \textbf{1.00$\pm$0.00} & 0.02$\pm$0.01 & 0.24$\pm$0.08 & \textbf{0.43$\pm$0.11}& 0.96$\pm$0.02\\
\toprule
Kaggle & 0.09$\pm$0.18&   0.09$\pm$0.17&   0.12$\pm$0.21 &  0.20$\pm$0.31 &  0.25$\pm$0.31& 0.44$\pm$0.37 &  0.40$\pm$0.41&  0.64$\pm$0.38 &  0.33$\pm$0.34 & 0.63$\pm$0.36 &  \textbf{0.64$\pm$0.34}& \textbf{0.93$\pm$0.10}\\
\toprule
Average &    0.22$\pm$0.31&   0.22$\pm$0.31&   0.23$\pm$0.32&  0.27$\pm$0.34&  0.33$\pm$0.34&  0.41$\pm$0.35 &  0.47$\pm$0.37&       0.65$\pm$0.31&    0.38$\pm$0.35&    0.61$\pm$0.35 & \textbf{0.57$\pm$0.37} & \textbf{0.90$\pm$0.21} \\
\end{tabular}
\end{table*}
In Table~\ref{tab:dataset_runtime}, we highlight the best performing approach of each row and hash size in \textbf{bold} font.
We observe that \system with \hash outperforms all the other baselines on all of the queries.
%\hash leverages simple syntactic information of the cell values to generate distinguishable index elements. 
\system + \hash can be up to $10x$ faster than \system + BF, which is the second most efficient baseline on average.
The cells showing this speedup are highlighted with \colorbox[HTML]{FFABA8}{red}.
BF's under-performance is rooted in its higher collision rate compared to \hash. Besides, by checking the length attribute that we use in \hash to generate the hash results, we can drop many of the non-joinable rows without evaluating all the bits in the super keys. This feature gives \system a superiority over BF in runtime even in the cases with similar or lower FP rates.
Another interesting insight is that \system + BF is faster than \system + HT in most of the cases and this is because BF is able to leverage more bits in the hash space to encode the join keys than a hash table that leverages only a single bit to hash each key value.%A single bit can only generate a limited number of unique hash values, e.g., 128 different hash values in 128-bit hash space, therefore, in the best case, $\frac{1}{128}$ of the single hashes will lead to collision, i.e., the same hash value. 
Table~\ref{tab:dataset_runtime} shows that \hash performs also better than LHBF in all of the cases. LHBF leverages fewer hash functions in comparison to BF with statistically similar FP probability. However, similar to HT, fewer hashing does not necessarily lead to a better performance in discovering the joinable tables. In our experiments LHBF only performs better on larger hash sizes for webtable queries, but in the remaining cases, LHBF is less efficient than BF.

The results also show that although super keys based on the standard hash functions, MD5, CityHash, SimHash, and Murmur, lead to overall performance gains over the naive SCR approach, all of them clearly fall behind \hash.
This is because they are not optimized to identify subset relationships that we encode via bitwise OR masking. They generate too many 1 bits (on average $50\%$) in the hash results, which leads to a higher number of FPs.
Thus, if a table contains six columns the aggregation of six hash results will on average turn $98\%$ of the super key to 1s, which would make the super key highly ineffective in masking.

%MD5, CityHash, SimHash, and Murmur lead to a very similar performance to SCR specially in open data and school experiments. This performance drop is because the row filtering overhead is almost as high as the runtime benefit from the row filtering.

Further, the experiments show that in most of the cases, a larger hash size (512 bits) leads to a faster join discovery than the smaller versions. However, in cases such as WT (100) and Simhash, the larger hash size results in slower discovery. The experiments that lead to slower runtime compared to their smaller hash size are marked as \colorbox[HTML]{A2EDFF}{blue} cells. 
When the FP rate is similar for two hash sizes, the approach with the larger hash size will result in similar or higher runtime. 
This is because the bit-wise OR operation is more expensive for larger hash values.
For instance, for WT (100) query tables and SimHash, the difference between the FP rate is almost zero, therefore, the runtime overhead of 512-bit hash causes worse performance.
In the more common case that a larger hash size reduces the number of FPs the resulting runtime is also lower.
For example, for the OD (100) datasets, the larger hash size for \hash results in $85\%$ fewer FPs on average, leading to $48\%$ speed-up.

Finally, we also observed that \system can indeed find interesting joinable tables based on composite keys. 
%Similarly, we observe that using composite keys we can find interesting tables for almost any dataset.  
% As another example, the most joinable table to the \textit{Page View} dataset~\cite{base_pageview} with query columns <``Name'', ``Country''>, which contains the information about Wikipedia pages of people, reveals the birth/death date of the people in the given countries. 
Searching for the top joinable tables to Kaggle Movie dataset based on a single-column join and ``Movie Title'' key columns, none of the top-10 tables contains more than one additional column, containing float numbers. However, using our multi-column join discovery approach and query columns <``Director name'', ``Movie Title''>, we obtain a table with 8 columns worth of information, including the plot of the movie, actor names, etc.
For the \textit{Kaggle Airline} dataset with join keys ``Airline name'' and ``Country'' \system obtains a table representing the airports in which the airlines operate flights in the given countries. 

\noindent\textbf{Summary.}
(i)~The syntactic feature extraction in \hash leads to a faster multi-attribute join discovery compared to the baselines;
(ii)~the standard hash functions are not ideal for n-ary joins due to their uniform distribution property that results in too many 1 bits; and
(iii)~ larger hash sizes result in higher calculation overhead.

\subsection{FP Rate}\label{subsubsection:fp}
%We now delve into more details regarding the FP rates.
There is a direct relationship between the table discovery runtime and the FP rate of the hash function used.
We measure the precision as the ratio of true positives (TP) over both TPs and FPs: $precision = \frac{TP}{TP + FP}$. Precision normalizes the TP-rate across datasets.

Table~\ref{tab:dataset_precision} shows the results of this experiment. Due to space reasons in this experiment, we only show the results for $128$- and $512$-bit hash sizes. Note that we observe the same trends as the other hash sizes.
Like runtime, precision increases with larger hash sizes in most of the cases. This is due to the fact that hash functions can scatter the values in a larger range and this leads to lower collision rate between super keys.
On average, \system + \hash achieves the highest precision compared to all the other approaches both in 128- and 512-bit hash spaces. \hash achieves up to $25\%$ higher precision than BF. 
% \hash maps the character and length features of the cell values to hash bits and rotates them based on the length. This removes the random overlap between the hashes. 
However, in five cases, BF outperforms \hash. In these experiments the difference between BF and \hash is only $4\%$.
\system's runtime superiority regardless of its precision is because of the design of \hash: the length feature of the cell value is located on the left-most segment of the array. Thus, the bit-wise OR operation can skip a table row before even getting to the character features of the cell values ensuring that this happens in the very first bit-segment of the hash. If the length of the key value has never appeared in the candidate row, \system moves to the next row.
In baseline approaches, such as BF, the 0-bits are randomly scattered. Therefore, step-wise filtering might take longer to find the contradicting bits.
% For query tables, such as WT (1k), \hash leads to distinctively high precision with 128 bits. 
%\hash achieves up to $25\%$ higher precision than BF. 
\hash displays the smallest standard deviation showing its robustness in achieving high precision.
In general, the precision achieved by LHBF is consistent with its runtime performance compared to the BF approach. LHBF achieves higher precision and ultimately, lower runtime for webtable queries with $512$-bit hash sizes but in the remaining experiments BF outperforms LHBF.
%As observed before, the SimHash function performs better than the other remaining hash functions. %CityHash performs very similarly to Murmur and MD5 except on WT (100), OD (100), and OD (10k), where CityHash gains up to $2\%$ more precision.
% As the precision experiments show, BF performs better than \hash in a single case of school query tables and $512$-bits hash size. However, due to the special hash design in \hash and the segmentation of the hash array, the evaluation process in \hash is much faster than BF. In \hash, a large portion of the FPs are pruned by only checking the length bits. This optimization prevents from checking all the remaining bits and leads to faster join discovery.

\noindent\textbf{Summary.}
(i) The precision of an approach correlates with its runtime.
(iii) The segmentation of bits in \hash leads to faster join discovery even in cases where precision rates are similar.

\subsection{\system In Depth}
\subsubsection{Top-$k$.}
The number of top-$k$ joinable tables changes the stopping criteria of the system (see table filtering step in Section~\ref{sec:index_applicaiton}).
Larger $k$ requires more tables to be evaluated.
In this experiment, we measure the precision of \system with different hash functions and vary $k$ from $2$ to $20$.
We ran the experiment with the WT (100) datasets as the input dataset against the DWTC corpus.
Other variables, such as key columns, are fixed in this experiment.
%We limit this experiment to the best performing hash functions.
In this expeirment, \system + \hash achieves the highest precision in comparison to other approaches for all $k$ values.
Increasing $k$, the precision for \hash increases $4\%$, while the precision remains the same for BF. The other hash functions lead to a slight precision cut encountering new tables with lower candidate joinable rows. This shows that \hash has higher ability to filter non-related rows compared to the given baselines. According to our observations on the candidate tables, this variation in precision occurs when the candidate tables contain more columns than average.
%Because of the reason that \hash is able to encode the cell values depending on their syntactical differences, the values from different domains map to different hash functions and these achieves fewer FPs.

% Figure~\ref{fig:topk} shows the results. \system + \hash achieves the highest precision in comparison to other approaches for all $k$ values.
% As expected, increasing $k$, increases the number of validated PL items because the first rule in the table filtering step starts later. Therefore, the system evaluates more candidate tables to discover the top-$k$ joinable tables. 
% According to the experiment, increasing $k$, the precision for \system increases, where, generally, the other hash functions lead to a slight precision cut encountering new tables with lower candidate joinable rows. This shows that \hash has higher ability to filter non-related rows compared to the given baselines. According to our observations on the candidate tables, this variation in precision occurs when the candidate tables contain more columns than average.
% Because of the reason that \hash minimizes the number of used bits in the candidate rows, the super key does not lose its ability in filtering irrelevant rows even in the existence of high dimensional tables. 
% This is not the case for BF. Because of the higher FP rate, more tables and in particular tables with large dimensions are passed to BF, which BF fails to filter.

\subsubsection{\hash components}
\begin{figure}[t!]
% \vspace{-.2cm}
\centering
\begin{tikzpicture}[scale=0.7]
  \begin{groupplot}[label style={font=\LARGE},
    	tick label style={font=\LARGE},
% 		xlabel={Approaches},
% 		ylabel={FP},
% 		label style={font=\large},
% 		tick style={font=\large},
		legend style={legend columns=4,at={(0.5,-0.25)},anchor=north,font=\Large},
	    x label style={at={(axis description cs:0.5,-0.1)},anchor=north},
		symbolic x coords = {Approaches},
		x tick label style  = {text width=1cm,align=center},
		xtick=data,
		ybar,
% 		x=5.0cm,
        width = 8cm,
        height = 5cm,
		ymin=0.0,
% 		ymode = log,
		ymax=1.2,
		ytick={0.0, .2, .4, .6, .8, 1},
		xmajorgrids=true,
		ymajorgrids=true,
		grid style=dashed ]
    \nextgroupplot [ylabel={Precision}]

%         \addplot[fill=red] coordinates{(Approaches, 0.14)}; \addlegendentry{SCI}
% 		\addplot[fill=purple] coordinates{(Approaches,0.21)}; \addlegendentry{Length}
% 		\addplot[fill=olive] coordinates{(Approaches,0.31)}; \addlegendentry{Rare characters}
% 		\addplot[fill=teal] coordinates{(Approaches,0.58)}; \addlegendentry{Char. + loc.}
% 		\addplot[fill=violet] coordinates{(Approaches,0.63)}; \addlegendentry{Char. + len. + loc.}
% 		\addplot[fill=cyan] coordinates{(Approaches,0.65)}; \addlegendentry{\hash (128 bit)}
% 		\addplot[fill=blue] coordinates{(Approaches,0.89)}; \addlegendentry{\hash (512 bit)}
% 		\addplot[fill=orange] coordinates{(Approaches,1.0)}; \addlegendentry{Ideal system}

        \addplot[fill=red] coordinates{(Approaches, 0.14)}; \addlegendentry{SCI}
		\addplot[fill=purple] coordinates{(Approaches,0.19)}; \addlegendentry{Length}
		\addplot[fill=olive] coordinates{(Approaches,0.28)}; \addlegendentry{Rare characters}
		\addplot[fill=teal] coordinates{(Approaches,0.54)}; \addlegendentry{Char. + loc.}
		\addplot[fill=violet] coordinates{(Approaches,0.59)}; \addlegendentry{Char. + len. + loc.}
		\addplot[fill=cyan] coordinates{(Approaches,0.61)}; \addlegendentry{\hash (128 bit)}
		\addplot[fill=blue] coordinates{(Approaches,0.88)}; \addlegendentry{\hash (512 bit)}
		\addplot[fill=orange] coordinates{(Approaches,1.0)}; \addlegendentry{Ideal system}
    
  \end{groupplot}
\end{tikzpicture}
% \vspace{-.2cm}
\caption{The influence of \hash components on Precision.}
\label{fig:components}
%\vspace{-.2cm}
\end{figure}
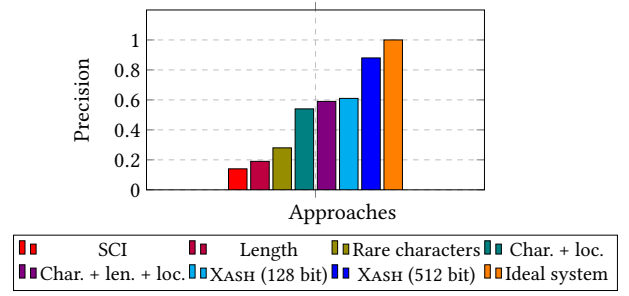
We now evaluate the impact of each feature used in \hash on the average precision and the FP rate of \system.
We use the WT (100) datasets in this experiment with the same setup as in the previous experiments.
According to the results in Figure~\ref{fig:components}, encoding the \textit{character and its location} has higher filtering power than the \textit{length} feature.
The difference between \hash and \textit{character + length + location} is the rotation operation.
In particular, we observe that the rotation filters $20\%$ of the remaining FPs undiscovered in \textit{character + length + location}.
This shows that the rotation plays an important role in pruning FPs.

\subsubsection{Join-Key Size}
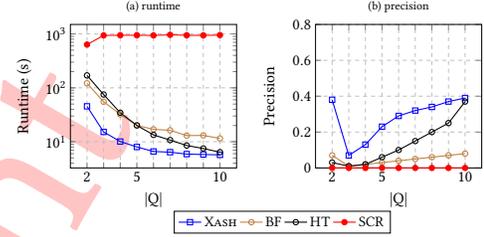
\begin{figure}[t!]
% \vspace{-.2cm}
\begin{tikzpicture}[scale=0.5]
  \begin{groupplot}[xtick=data, group style={group size=2 by 1 , horizontal sep=1.0cm}, width = 10cm, every node near coord/.append ,nodes near coords, point meta=explicit symbolic, log origin=infty]%style={yshift=-0.21cm}
    % baseline
    % TR
    % AugX
    
    \nextgroupplot [title = (a) runtime,
    	xlabel={|Q|},
    	ylabel={Runtime (s)},
    	label style={font=\LARGE},
    	tick label style={font=\LARGE},
    	legend style={legend columns=2,at={(0.5,-0.2)},anchor=north,font=\Large},
    	width = 6cm,
    	height = 5.4cm,
    	xmin=1,
    	xmax=11,
    	ymin= 0,
    	ymax=1500,
    % 	log basis x={2},
    % 	log basis y={10}, 
    % 	xticklabels={1, , 5,,,,15,,,},
        every axis plot/.append style={thick},
    	ymode=log,
    	xtick={2, 3, 4, 5, 6, 7, 8, 9, 10},
    	ytick={0, 1, 10, 100, 1000},
    	xticklabels={2,,,5,,,,,10,,,,,15},
    	xmajorgrids=true,
    	ymajorgrids=true,
    	grid style=dashed]
    	\addplot[color=blue,mark=square] coordinates{(2, 45.4)(3, 15.2)(4, 10.1)(5, 8)(6, 6.6)(7, 6.4)(8, 5.9)(9, 5.8)(10, 5.7)};% \addlegendentry{\hash}
    	\addplot[color=brown,mark=o] coordinates{(2, 121.4)(3, 55.2)(4, 32.2)(5, 20.1)(6, 16.97)(7, 16.3)(8, 13)(9, 13.1)(10, 11.52)}; %\addlegendentry{BF}
    	\addplot[color=black,mark=o] coordinates{(2, 170.1)(3, 75.2)(4, 34.5)(5, 20.1)(6, 13.4)(7, 10.74)(8, 8.52)(9, 7.5)(10, 6.4)}; %\addlegendentry{HT}
        \addplot[color=red,mark=*] coordinates{(2, 634.8)(3, 938.5)(4, 945.9)(5, 946.1)(6, 938.6)(7, 962.3)(8, 943.1)(9, 939.9)(10, 952.1)}; %\addlegendentry{DXF}

    \nextgroupplot [title = (b) precision,
    style={xshift=1.1cm},
    	xlabel={|Q|},
    	ylabel={Precision},
    	label style={font=\LARGE},
    	tick label style={font=\LARGE},
    	legend style={legend columns=5,at={(-0.2,-0.3)},anchor=north,font=\Large},
    	width = 6cm,
    	height = 5.4cm,
    	xmin=1,
    	xmax=11,
    	ymin= 0,
    	ymax=.8,
    % 	log basis x={2},
    % 	log basis y={10}, 
    % 	xticklabels={1, , 5,,,,15,,,},
        every axis plot/.append style={thick},
    % 	ymode=log,
    	xtick={2, 3, 4, 5, 6, 7, 8, 9, 10},
    	ytick={0, .2, .4, .6, .8, 1},
    	xticklabels={2,,,5,,,,,10,,,,,15},
    	xmajorgrids=true,
    	ymajorgrids=true,
    	grid style=dashed]
    % 	\addplot[color=orange,mark=x] coordinates{(2, 6680)(3, 221)(4, 217)(5, 217)(6, 217)(7, 217)(8, 217)(9, 217)(10, 217)}; \addlegendentry{True positives}
    % 	\addplot[color=blue,mark=square] coordinates{(2, 17587)(3, 3097)(4, 1618)(5, 955)(6, 741)(7, 677)(8, 634)(9, 582)(10, 563)}; \addlegendentry{\hash}
    % 	\addplot[color=brown,mark=o] coordinates{(2, 92883)(3, 23341)(4, 10699)(5, 6440)(6, 5017)(7, 3988)(8, 3483)(9, 3142)(10, 2854)}; \addlegendentry{BF}
    % 	\addplot[color=black,mark=o] coordinates{(2, 211800)(3, 33132)(4, 9645)(5, 3561)(6, 2062)(7, 1405)(8, 1062)(9, 868)(10, 586)}; \addlegendentry{HT}
    %     \addplot[color=red,mark=*] coordinates{(2, 1958951)(3, 1958951)(4, 1958951)(5, 1958951)(6, 1958951)(7, 1958951)(8, 1958951)(9, 1958951)(10, 1958951)}; \addlegendentry{DXF}
        
    	\addplot[color=blue,mark=square] coordinates{(2, .38)(3, .07)(4, .13)(5, .23)(6, .29)(7, .32)(8, .34)(9, .37)(10, .39)}; \addlegendentry{\hash}
    	\addplot[color=brown,mark=o] coordinates{(2, .07)(3, .01)(4, .02)(5, .03)(6, .04)(7, .05)(8, .06)(9, .07)(10, .08)}; \addlegendentry{BF}
    	\addplot[color=black,mark=o] coordinates{(2, .03)(3, .01)(4, .02)(5, .06)(6, .1)(7, .15)(8, .20)(9, .25)(10, .37)}; \addlegendentry{HT}
        \addplot[color=red,mark=*] coordinates{(2, 0)(3, 0)(4, 0)(5, 0)(6, 0)(7, 0)(8, 0)(9, 0)(10, 0)}; \addlegendentry{SCR}
		
  \end{groupplot}
\end{tikzpicture}
% \vspace{-.2cm}
\caption{Key size experiment.}
\label{fig:keysize}
\end{figure}
Here, we evaluate the scalability of \system in the existence of different join-key sizes, i.e.,~the number of columns in the composite key of the input table. 
Due to the limited number of tables with high dimensional composite key, we ran this experiment with a random dataset from the German Open Data corpus with up to 10 columns that can form a composite key (out of 33 columns).
Figure~\ref{fig:keysize} (a) and \ref{fig:keysize} (b) depict the runtime and precision results for different key sizes.
Increasing the number of columns, we observe that the runtime for \system constantly reduces because the FP rate constantly decreases. However, this does not imply a constant increase in precision because increasing the number of key columns changes the ratio of joinable and non-joinable rows. In our experiment, moving from a composite key with two columns to a three-column key, $97\%$ of the joinable rows become non-joinable due to the newly introduced key column. Thus, the filter is confronted with a larger number of FPs, some of which come through.
From key size $4$ upwards, precision increases again as expected.
The runtime gain for larger key sizes is because of two reasons:
\textit{(1)}  With more columns in the query join key, there will be more 1-bits in the query super key, which makes it harder to mask. 
\textit{(2)} Increasing the size of the composite key leads to fewer joinable table rows. Therefore, it is more likely that the second table filtering rule drops the candidate tables without evaluating the remaining rows.
%Again, \hash outperforms other baselines in runtime and precision.   

\subsubsection{Initial column selection.}
We evaluate our heuristic to select the initial query column by comparing the number of retrieved PL items in \system to four other baselines:
\textit{(i)} The column order. In this approach, the initial column will be the first query column according to the column order inside the table. %The idea behind this strategy is that usually, the identifying columns come earlier in the table.
\textit{(ii)} The longest string (TLS). In this approach, the system selects the column that contains the longest cell value as the initial column. %The heuristic here is that the longer the text, the more specific is the column. 
\textit{(iii)} The worst-case scenario. A hypothetical approach that always picks the worst column that returns the larger number of PL items.
\textit{(iv)} Best, i.e. ground truth, which chooses the column that filters most.
The best- and the worst-case scenarios provide lower and upper bounds for our experiment.
This experiment is done using the OD (10k) tables.
In this experiment our cardinality-based heuristic outperfmed the other heuristics. It retrieved on average only 179 PLs compared to Column Order, TLS, and the worst-case scenario with 202, 248, and 728 Pls, respectively. The optimal number of PLs based on ground truth was 83. The heuristic used in \system performs better because of the fact that the number of PL items per cell value follows the power-law distribution. There is a small set of values that have a large number of PL items but most of the values lead to a similar number of PL items (The average number is $12$).
%Table~\ref{tab:ICS} shows that the cardinality-based heuristic used in \system, performs best and is very close to the ground truth approach. The heuristic used in \system performs better because of the fact that the number of PL items per cell value follows the power-law distribution. There is a small set of values that have a large number of PL items but most of the values lead to a similar number of PL items (The average number is $12$).% Thus, there is a direct relationship between the cardinality and the number of retrieved PL items.

\section{Related Work}\label{sec:related}
N-ary joinable table search is not a well-researched area and a very limited number of works have focused on multi-attribute joins discovery. Li et al.'s index design~\cite{li2015efficient} is one of these works. They introduce a prefix tree index to calculate the joinability score between two given tables for n-ary joins. 
The drawback of this approach is that it is not scalable to data lakes. The prefix tree-based approach assumes that the one-to-one mapping between the composite key columns and the columns in the candidate tables is apriori known. 
\system, on the other hand, leverages an index structure that does not require the user to provide the mappings. 

There is a large number of research papers that address joinable tables discovery problem based on unary keys~\cite{zhu2019josie, xiao2009top, eberius2015top, das2012finding, yakout2012infogather, zhang2013infogather+, fernandez2018aurum, DBLP:journals/pvldb/ZhuNPM16, DBLP:conf/sigmod/ZhangI20}. 
\system extends the standard single-attribute inverted index used in these state-of-the-art systems and applies a fast single-operation filtering approach to be able to detect joinable table rows without actual value comparison. Any single attribute inverted index can be extended with the super key used in \system.

For spatial indices, the goal of an ideal n-ary join discovery is to map multiple dimensions into an easy to search index. However, spatial index structures are a very special case compared to the problem we tackle in this paper. The spatial indices e.g. KD tree~\cite{bentley1975multidimensional}, KDB tree~\cite{robinson1981kdb}, R+ tree~\cite{sellis1987r+}, Geohash~\cite{geohash}, grid files~\cite{nievergelt1984grid}, suffer from three drawbacks: 1) the indexed data points always have fixed dimensions, such as latitude and longitude. 2) These dimensions have a fixed order. For instance, latitude is always compared to the latitude of the other points. In other words, there is a pre-defined mapping between the dimensions. However, in join discovery, key values can appear in arbitrary columns and comparing the right values is not a straightforward task. 3) These indices are designed to work with numerical values. 
\system on the other hand does not have any of the aforementioned assumptions. It discovers the joinable table rows with an arbitrary number of columns. Also, \system leverages \hash and discovers the key values regardless of their position in the candidate rows. Besides, \hash is applicable for numerical and alphabetical column values.

Data enrichment solutions attempt to join additional tables to improve the downstream machine learning accuracy. Data enrichment solutions also either require pre-defined mappings between joinable columns~\cite{chepurko13arda, kumar2016join}, which needs the user input on every single external table, or they focus on single column joins~\cite{zhu2019josie, xiao2009top, eberius2015top, das2012finding, esmailoghli2021cocoa}. \system calculates the joinability of the candidate tables without any apriori knowledge, e.g. mapping information. This allows \system to scale to any number of external tables.

Inclusion dependency (IND) discovery~\cite{de2009unary, papenbrock2015divide, koeller2003discovery} is a research direction where the goal is to discover foreign keys in the database~\cite{papenbrock2015divide}. The focus is on identifying fitting column combinations across all datasets. In data discovery, we are interested in the top-k tables with the most fitting INDs.
\section{Conclusion and Future Work} \label{sec:conclusion}
In this paper, we tackled the problem of n-ary or multi-attribute join search. We proposed a hash-based filtering approach \system that removes redundant table rows from further joinability calculation processes to increase the scalability of the join discovery. We proposed \hash, a hash function that leverages syntactic features of the cell values that ultimately lead to the lowest FP rates. We showed that \system is more effective and efficient than state-of-the-art systems.
We focused on the equi-joins. Because \hash uses syntactic features including the character and length features of the cell values, it has the potential to discover similarity joins as well. According to our observations, the false positives caused by \hash were those that are syntactically similar to the actual key values, e.g., discovering the composite key <``brooklyn'', ``cambridge''> instead of <``brooklyn'', ``bay ridge''>. 
Another future direction is to reason about hashing short key values. \hash cannot use its optimal potential if cell values are too short.

\begin{acks} 
This project has been supported by the German Research Foundation (DFG) under grant agreement 387872445 and the German Ministry for Education and Research as BIFOLD — “Berlin Institute for the Foundations of Learning and Data” (01IS18025A and 01IS18037A).
\end{acks}

\balance

% \small
%\bibliographystyle{abbrv}
 \bibliographystyle{ACM-Reference-Format}
\bibliography{references}

\end{document}